\documentclass[useAMS,usenatbib]{mn2e}
\usepackage{graphicx,amsmath,txfonts}

\title{On the evolution of vortices in massive protoplanetary discs}
\author[Arnaud Pierens, Min-Kai Lin]{Arnaud Pierens $^{1,2}$, Min-Kai Lin $^{3}$ \\
$^1$Universit\'e de Bordeaux, Observatoire Aquitain des Sciences de l'Univers,
    BP89 33271 Floirac Cedex, France \\
$^2$CNRS, Laboratoire d'Astrophysique de Bordeaux,
     BP89 33271 Floirac Cedex, France\\
 $^3$ Institute of Astronomy and Astrophysics, Academia Sinica,
Taipei 10617, Taiwan}
\date{Released 2012 Xxxxx XX}

\pagerange{\pageref{firstpage}--\pageref{lastpage}} \pubyear{2014}

\def\LaTeX{L\kern-.36em\raise.3ex\hbox{a}\kern-.15em
    T\kern-.1667em\lower.7ex\hbox{E}\kern-.125emX}
\begin{document}
\label{firstpage}
\maketitle
\begin{abstract}
 It  is expected that a pressure bump can be formed at the inner edge of a dead-zone, and where vortices can develop through the Rossby Wave Instability (RWI). It has been suggested that self-gravity can significantly affect the evolution of such vortices.   
We present the results of 2D  hydrodynamical simulations of the evolution of vortices forming at a pressure bump in self-gravitating discs with Toomre parameter in the range $4-30$. We consider isothermal plus non-isothermal disc models  that employ either the classical $\beta$ prescription  or a more realistic treatment for cooling.  The main aim is to investigate whether the condensating effect of self-gravity can stabilize vortices in sufficiently massive discs. 
We confirm that in isothermal disc models with ${\cal Q} \gtrsim 15$, vortex decay occurs due to the vortex self-gravitational torque. For discs with $3\lesssim {\cal Q} \lesssim 7$,  the vortex  develops gravitational instabilities within its core and undergoes  gravitational collapse,  whereas more massive discs give rise to the formation of global eccentric modes.   In  non-isothermal discs with $\beta$ cooling,  the vortex maintains a turbulent core prior to undergoing gravitational collapse for  $\beta \lesssim 0.1$,  whereas it decays if $\beta \ge 1$. 
  In  models that incorpore both self-gravity and a better treatment for cooling, however, a stable vortex is formed  with aspect ratio $\chi \sim 3-4$.
Our results indicate that self-gravity significantly impacts the evolution of vortices forming in protoplanetary discs, although the  thermodynamical structure of the vortex  is equally important for determining its long-term dynamics.

\end{abstract}

\section{Introduction}

A striking feature of the current population of exoplanets is the broad diversity in system architectures that have been discovered. Among the known multiplanetary systems, a significant number consist of compact and non-resonant systems of Super-Earths and mini-Neptunes orbiting within a few tenths of AU from their stars (e.g. Lissauer et al. 2011; Lovis et al. 2011). There are two basic scenarios for the formation of these systems. The first one corresponds to the {\it in-situ} formation of these systems, involving planetesimal accretion within high-mass discs (Hansen \& Murray 2012, Chiang \& Laughlin 2013). This model however requires extremely massive protoplanetary discs with a broad range of surface density slopes, which is inconsistent with any known disc theory (Raymond \& Cossou 2013). An alternative possibility is that systems of close-in Super-Earths formed through accretion during the inward migration of planetary embryos (Terquem \& Papaloizou 2007; Ogihara \& Ida 2009; Cossou et al. 2014). Super-Earths formed in this way are expected to migrate to the inner edge of the disc where they pile up in long chains of resonances. Resonant configurations can  subsequently be broken during disk dissipation (Izidoro et al. 2017),  or due to  the action of disc turbulence (Pierens et al. 2011; Rein 2012) or interaction with planet wakes (Baruteau \& Papaloizou 2013).\\
In the context of this model, understanding the physical conditions at the inner edge of the disc is therefore a crucial issue. In the very inner  regions where temperatures are high enough for the disc to be thermally ionized, turbulence is expected to be driven by the magnetorotational instability (MRI) (Balbus \& Hawley 1991). Beyond  $\sim 0.8$ AU typically (Flock et al. 2017), however, temperature drop below the ionization threshold of alkali metals ($T\sim 1000$ K) prevents turbulence to be sustained, giving rise to the formation of a dead-zone (Gammie 1996). Since magnetic stresses are order of magnitude weaker there,  gas accretion is driven faster in the active region than in the dead-zone. This results in the formation of a local maximum in the radial profile of the disc pressure,  where  dust particles can be collected  and which can  act as a planet trap (Masset et al. 2006). At this location, the recent 3D radiation MHD simulations of Flock et al. (2017) have also demonstrated that for realistic inner disc models, a vortex is likely to be formed  through the Rossby Wave Instability (RWI) (Lovelace et al. 1999). Vortex formation through the RWI has also been invoked as a possible origin for the non-axisymmetric structures that have been observed in many transition discs, like in Oph IRS 48 (van der Marel et al. 2013), or HD 142527 (Casassus et al. 2013; Fukagawa et al. 2013). In that case, the pressure maximum responsible for driving the RWI   should be rather located at the outer edge of the dead-zone  where the disc transitions from low to high gas ionisation where the MRI is active (Flock et al. 2017),  or located at the edge of a planetary gap (Ataiee et al. 2013; Zhu et al. 2014).\\
Because of the reduction of the accretion stress within the dead-zone, a significant amount of mass can accumulate there such that the disc self-gravity starts to affect the structure and evolution of vortices that are produced by the RWI.
Results from previous work indicate that  self-gravity tends to stabilize vortex forming instabilities. Linear stability analysis (Lovelace \& Hohlfed 2013) predicts that self-gravity can inhibit modes with azimuthal wavenumbers $m<(\pi/2)/(hQ)$, where $h$ is the disc aspect ratio and ${\cal Q}$ the Toomre stability parameter.  In presence of self-gravity, suppression of modes with low $m$ values has also been observed both in 2D (Lin \& Papaloizou 2011) and 3D (Lin 2012) simulations of  vortices forming at the edge of the gap opened by a giant planet.  For a disc which initially presents a density bump where the RWI can grow,  this has also been in reported in the simulations of Zhu \& Baruteau (2016), who also found that large-scale vortices are significantly weakened when self-gravity is included. In fact, it is expected that self-gravity can affect vortex dynamics in discs with ${\cal Q}\lesssim 1/h$ (Regaly et al. 2012; Yellin-Bergovoy et al. 2015), which suggests a possible important role of self-gravity on the RWI for moderatly massive discs. This has been recently confimed by  Regaly \& Vorobyov (2017) who have shown that vortices developed at sharp viscosity transitions can be significantly stretched by the effect of self-gravity in low-mass discs with masses $0.001 \le M_{discs}/M_\star \le 0.01$, where $M_\star$ is the mass of the central star.  This basically arises because the vortex exerts a negative (resp. positive) torque on the gas material located ahead (resp. behind) of the vortex which, combined with the effect of Keplerian shear, results in the vortex being significantly stretched in the azimuthal direction.  As mentionned by Regaly \& Vorobyov (2017), however, it is not clear whether or not vortex stretching can overcome the effect of self-gravitational contraction in more massive discs. \\
In this paper we examine by means of 2D hydrodynamical simulations the effect of self-gravity on the evolution of vortices that are formed at the inner edge of the dead-zone in massive discs with  ${\cal Q}_{init}\le 30$, where ${\cal Q}_{init}$ is the initial value for the Toomre parameter at the location of the viscosity transition. The aim is to test whether gravitational condensation can eventually stabilize the vortex structure in massive discs, and how this depends on the equation of state that is adopted.  The paper is organized as follows. In Section $2$ we describe the hydrodynamical model and the initial conditions that are used in the simulations.  In Section $3$ we discuss the effect of self-gravity on the evolution of vortices in  both  isothermal discs while we consider the case of non-isothermal discs in Section $4$.   Finally, we discuss our results and draw our conclusions in Section $5$.

\section{The hydrodynamical model}
\subsection{Numerical setup}

\begin{figure*}
\centering
\includegraphics[width=\textwidth]{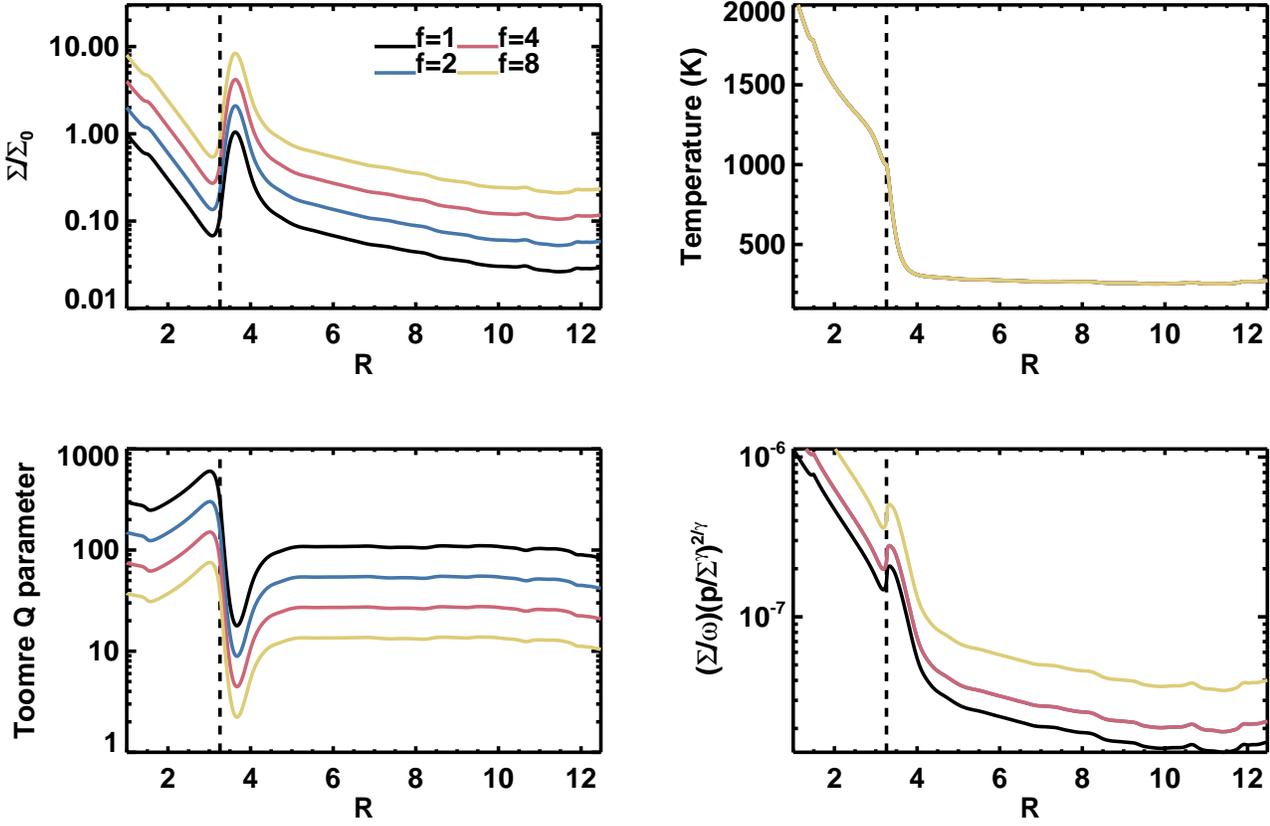}
\caption{Radial profiles of the surface density (upper left panel), temperature (upper right panel),  Toomre parameter (lower left panel) and ${\cal L}$ function defined as ${\cal L}=(\Sigma/\omega)(p/\Sigma^\gamma)^{2/\gamma}$. These profiles  correspond to the initial conditions in radiative disc models for $f=1, 2, 4, 8$.  The vertical dashed line shows the location of the viscosity transition.}
\label{fig:init}
\end{figure*}

Simulations were performed using the GENESIS (De Val-Borro et al. 2006) numerical code which solves 
the equations governing the disc evolution on a polar grid $(R,\phi)$ using an advection scheme based on the monotonic  transport 
algorithm (Van Leer 1977). It uses the FARGO algorithm (Masset 2000) to avoid time step limitation due to the 
Keplerian velocity at the inner edge of the disc, and includes a module to calculate self-gravity using a Fast Fourier Transform (FFT) method (Baruteau \& Masset 2008).  In order to take into account the effect of the finite disc thickness, the gravitational potential is smoothed out using a softenting length $\epsilon_{SG}=b R$ with $b=0.6h$ (Muller \& Kley 2012; Zhu \& Baruteau 2016).  We emphasize that the indirect term of the gravitational potential has been included in the simulations presented here. Previous work has demonstrated that including this term tends to  strenghen large-scale vortices (Zhu \& Baruteau 2016; Regaly \& Vorobyov 2017,  so that it is important to include this term when considering the evolution of vortices formed at a viscosity transition.\\
We adopt computational units such that the mass of the central star is $M_*=1M_\odot$, the gravitational constant is 
$G=1$, and the radius $R=1$ in the computational domain corresponds to $0.3$ AU.  When discussing the results of the simulations, 
 time will be measured in units of  the orbital period at the $R=3.5$, which corresponds to the location of the pressure bump where the RWI is generated. \\
The simulations presented here employ $N_R=792$ radial grid cells logarithmically distributed between $R_{in}=1$ and 
$R_{out}=12.5$, and $N\phi=1280$ grid cells in azimuth. \\
 In this work, we will examine the evolution of vortices in isothermal and radiative disc models.  In both cases,  the effective kinematic viscosity is modeled using the standard $\alpha$ prescription $\nu=\alpha c_s^2/\Omega$ (Shakura \& Sunyaev 1973), where $c_s$ is the isothermal sound speed,  $\Omega$ the 
angular velocity, and $\alpha$  is the viscous stress parameter that can vary from very small values in the dead zone to much higher values in the 
active region.
\subsubsection{Isothermal disc models}

 For the  isothermal disc models, the $\alpha$ profile that we employ is given by:
\begin{equation}
\alpha=\frac{\alpha_a-\alpha_d}{2}\left(\tanh\left(\frac{R-R_{idz}}{\Delta R_{idz}}\right)-1\right)+\alpha_a
\label{eq:alpha-iso}
\end{equation}

where $\alpha_a=10^{-2}$ is the $\alpha$ value  in the active region, whereas the $\alpha$ viscosity inside the dead zone is set 
to $\alpha_{dz}=10^{-4}$. We note that adopting a non-zero value for $\alpha_{dz}$ is in agreement with the results of 3D magnetohydrodynamical simulations (Okuzumi \& Hirose 2011; Gressel et al. 2012) which show that the dead zone has a small residual viscosity due to the propagation, inside the dead zone, of sound waves generated in the active region. In 
the previous equation $R_{idz}=3.5$ corresponds to the location of the inner edge of the dead zone and $\Delta R_{idz}$ is the radial width of the viscosity transition which is 
set to $\Delta R_{idz}=H_{idz}$, where $H_{idz}$ is the disc scale height at $R=R_{idz}$. \\

\subsubsection{Non-isothermal and radiative disc models}

The energy equation that is implemented in the code for the radiative disc models reads:
\begin{equation}
\frac{\partial e}{\partial t}+\nabla \cdot (e{\bf v})=-(\gamma-1)e{\nabla \cdot {\bf v}}+{\cal Q}^+_{visc}-{\cal Q}^-\label{eq:energy}
\end{equation}
where $e$ is the thermal energy density, $\bf v$ the velocity, $\gamma$ the adiabatic index which is set 
to $\gamma=1.4$. In the previous equation, ${\cal Q}^+_{visc}$ is the viscous heating term ( e.g. D'angelo et al. 2003) and ${\cal Q}^-$ is the gas cooling function for which we adopt two different forms. The first one corresponds to a standard $\beta$ parametrization for cooling, with (Les \& Lin 2015):
\begin{equation}
{\cal Q}^-=\frac{1}{\tau_{cool}}\left(e-e_i\frac{\Sigma}{\Sigma_i}\right)
\label{eq:cool}
\end{equation}
 where $e_i$ and $\Sigma_i$ are the initial thermal energy and surface density which are set in such a way that both the aspect ratio and surface density profiles coincide to those of the isothermal runs just before the random perturbations that will give rise to vortices are applied (see Sect. \ref{sec:init}). $\tau_{cool}$ is the cooling time which is set to $\tau_{cool}=\beta \Omega^{-1}$, with $\beta$ the cooling parameter. In this work, we considered values for $\beta$ 
running from $\beta=0.01$ to $\beta=1$.  We emphasize that in the context of gravitational instabilities, a disc that becomes gravitationally unstable can fragment for these  values for $\beta$ (Gammie 2001).  Here , the disc  is initially stable regarding the development of gravitational instabilities since ${\cal Q}_{init}\gtrsim 2$ at all radii and for all models that we consider. Whenever the cooling term given by Eq. \ref{eq:cool} is employed, we also note that  an additional source term is also included in the energy equation by changing ${\cal Q}^+_{visc}$ to 
${\cal Q}^+_{visc}-{\cal Q}^+_{visc,i}\frac{\Sigma}{\Sigma_i}$ to ensure that the heating and cooling terms couterbalance initially, where the index $i$ denotes the initial evaluation of the quantities ${\cal Q}, \Sigma$.\\

We also considered models with more realistic cooling instead of $\beta$ cooling. In that case the cooling function is similar to that used in Faure et al. (2015) and is  given by:

\begin{equation}
{\cal Q}^-=b \Sigma (T^4-T_{irr}^4)
\label{eq:cooling}
\end{equation}
 where $T=\mu(\gamma-1)e/{\cal R}\Sigma$ is the temperature, with ${\cal R}$  the gas constant and  $\mu=2.3$ the mean molecular weight. $b$ is the cooling parameter and  $T_{irr}$ is the irradiation temperature with:
\begin{equation}
T_{irr}^4=0.1\left(\frac{R_\star}{R}\right)^2T_\star^4
\end{equation}
where $T_\star=4300 K$ is the stellar temperature,  $R_\star=2 R_\odot$ the stellar radius, and  where the factor of $0.1$ accounts for the component of stellar irradiation   
that is normal to the disk equatorial plane (Zhu et al. 2012).  A constant value is  adopted for the cooling parameter $b$  in Eq. \ref{eq:cooling}, and is chosen in such a way that the disc aspect ratio $h$ 
at the inner edge of the disc corresponds to a chosen value $h=h_{in}$.  Given that ${\cal Q}^+_{visc}=\frac{9}{4}\nu \Sigma \Omega^2$ and 
${\cal Q}^-\sim b \Sigma T^4$ in the active region, the condition of thermal equilibrium implies:

\begin{equation}
b=\frac{9 {\cal R}^4 \alpha_a}{\mu ^4h_{in}^6 R_{in}^6\Omega_{in}^5}
\end{equation}
where  $\Omega_{in}$ the angular velocity at $R=R_{in}$.\\\\

To allow for a direct comparison with the  results of Faure et al. (2015), the  runs with realistic cooling include a different 
function for $\alpha$:
\begin{equation}
\alpha=
\begin{cases} 
\alpha_a &  \text{if} \; T \ge T_{MRI}\\
\alpha_{dz}  & \text{if} \; T <T_{MRI}
\end{cases}
\label{eq:alpha}
\end{equation}

where $T_{MRI}=1000$ K is the critical temperature above which the MRI is supposed to be at work due to thermal ionization of alkali metals (Umebayashi \& Nakano 1988).  It is important to note that the results for  such runs may not be easily compared to those corresponding to an non-isothermal setup with $\beta$ cooling. This is because i) Eq. \ref{eq:alpha}  employs a feedback loop between the temperature and viscosity which is responsible for the vortex cycles observed in the simulations of Faure et al. (2015) and ii) compared to the case where $\alpha$ is given by Eq. \ref{eq:alpha-iso}, Eq. 
\ref{eq:alpha} gives rise to a much sharper viscosity transtion whose width corresponds typically to that of one grid cell,  and this  is expected to give rise to a stronger vortex  with longer lifetime (Regaly et al. 2012). \\

 In these runs, we also model the turbulent diffusion of  heat by adding in the right-hand side of Eq. \ref{eq:energy} a diffusion term for the gas entropy $S=p/\Sigma^\gamma$, where $p$ 
is the pressure, and which is given by ${\cal D}=\kappa e \nabla^2 \text{log}S$.  The thermal diffusion coefficient $\kappa$  is  chosen assuming a Prandtl number $P_R=\nu/\kappa$ of unity, 
which  is consistent with the results  of Pierens et al. (2012) who found $P_R\sim 1.2$ in non-isothermal disc models with turbulence driven by stochastic 
forcing. Within the dead zone, we note that radiative diffusion can possibly dominate over turbulent transport for diffusing 
entropy (Latter \& Balbus 2012), so we expect the Prandtl number in a realistic protoplanetary disc model to be close to unity only in the active region. 

\subsection{Initial conditions}
\label{sec:init}

\begin{figure*}
\centering
\includegraphics[width=\textwidth]{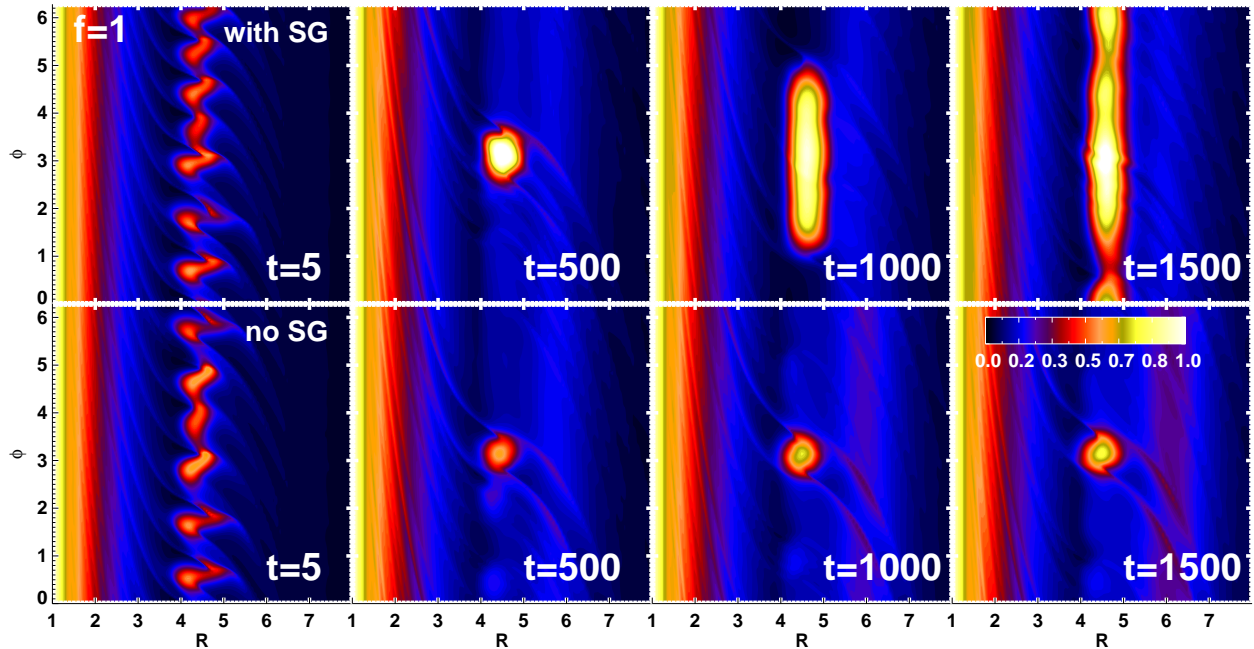}
\caption{{\it Upper panel:} contours of the scaled surface density $\Sigma/\Sigma_{in}$ at different times for the isothermal run with $f=1$ and with self-gravity included. {\it Lower panel:} same but in the case where self-gravity is not included. }
\label{fig:run1}
\end{figure*}

The  isothermal disc models that we consider have constant aspect ratio $h=0.05$, which corresponds to a fixed temperature profile that varies as $T\propto R^{-1}$. A similar initial temperature profile is used in the non-isothermal disc models. In the case where the $\beta$ cooling prescription is adopted, the cooling function given by Eq. \ref{eq:cooling} tries to restore the  initial temperature profile whereas in the case where a more realistic cooling is used, the temperature profile at steady-state can differ significantly from the original one. In both cases, however,    the aspect ratio at the disc inner edge  is kept fixed  with a value of $h=h_{in}=0.05$. \\
 The initial disc surface density is $\Sigma=f \Sigma_{in} (R/R_{in})^{-3/2}$ where $f$ is a scaling factor for which we consider values of $f=1,2,4,8$ and 
$\Sigma_{in}$ is defined such that for $f=1$,  the initial surface density at $1$ AU  in the unperturbed disc is $\sim 870$ $\text{g/cm}^2$. This corresponds to  a disc mass $M_d$ contained within the computational domain of $M_d\sim 0.002$ $M_\star$, whereas $M_d\sim 0.02$ for $f=8$.  \\
 To trigger the RWI within the disc, we proceed 
into two steps. We first make a pressure bump form and evolve as a result of  the radial viscosity transition given by Eq.  \ref{eq:alpha-iso} or \ref{eq:alpha}, until the amplitude of the density bump is five times 
the initial surface density.  In the upper left and upper right panels of Fig. \ref{fig:init}, we plot for the radiative case the resulting surface density  and  temperature  profiles at that time, which corresponds to $t\sim 380$ orbits at the location of the pressure bump. The lower left panel of Fig. \ref{fig:init} shows 
the Toomre $Q$ parameter,
$$
Q=\frac{\Omega c_s}{\pi G \Sigma}
$$
as a function of radius. As expected, the Toomre parameter is maximal in the active region due to the higher temperature there, and minimal at the location of the pressure bump, 
with values ranging from  $Q\sim 15$ for $f=1$ down to $Q\sim 2$ for $f=8$. Finally, the lower right panel of Fig. \ref{fig:init} 
displays the radial profile of the function:
\begin{equation}
{\cal L}=\frac{\Sigma}{\omega}\left(\frac{p}{\Sigma^\gamma}\right)^{2/\gamma}
\end{equation}
where $\omega$ is the vertical component of the vorticity.  According to linear theory (Lovelave et al. 1999; Li et al. 2000, 2001),  we expect the disc to be unstable to the RWI  at the location where the function ${\cal L}$ presents a maximum in its radial profile, namely at the location of the pressure bump from the lower right panel of Fig. \ref{fig:init}. This is also valid when an isothermal equation of state is adopted, with  the maximum in the ${\cal L}$ function corresponding to a minimum of the potential vorticity in that case.\\
The RWI is triggered by adding a $10^{-2}c_s$ amplitude white noise to the radial component of the velocity, which subsequently leads to the formation of vortices at the surface density maximum after
$\sim 10$ orbits.  

\subsection{boundary conditions}

 To avoid wave reflection at the edges of the computational domain, we employ damping boundary conditions (de Val-Borro et al. 2006) using wave killing zones for $R< 1.5$ and $R>12$ 
where the surface density, internal energy and velocity components are relaxed toward their values at the end of the first step (see Sect. \ref{sec:init}), namely prior to the addition of the random perturbation. As will be discussed later in the paper, interaction with a massive vortex can make the disc become globally eccentric. In that case, results from previous 
studies (Papaloizou 2005, Kley \& Dirksen 2006) suggest that the boundary conditions we adopt  are well suited to hydrodynamical simulations of eccentricity protoplanetary discs. We note that in test simulations employing an outflow boundary  condition at the inner edge, significant growth of the disc eccentricity was not observed, due to the overestimated 
loss of material and radial kinetic energy through the inner boundary, which is in  agreement with the results of Papaloizou (2005).

\subsection{Diagnostics}

To estimate the vortex strength, we calculate the Rossby number $Ro$ which is defined as the dimensionless vertical component of the vorticity relatively to the background flow:
\begin{equation}
Ro=\frac{{\bf e_z}\cdot(\nabla\wedge({\bf v}-R\Omega_K{\bf e_\phi}))}{2\Omega_K(R_v)}
\end{equation}
  Strength of the vortex can be estimated by measuring $Ro$ at vortex center. Regarding other vortex characteristics,  its  aspect ratio $\chi_v$ can be determined  by first locating the vortex boundary that we define as the contour where $\Sigma=0.5\max (\Sigma)$. Assuming that the shape of this contour is close to that of an ellipse, we can then estimate the vortex aspect ratio simply as $\chi_v=a/b$, where $a$ and $b$ are the semimajor and semiminor 
axes of the ellipse respectively. \\
Other diagnostic quantities include:
\begin{enumerate}
\item The reynolds stress $\alpha_R$ which is given by: 
\begin{equation}
\alpha_R=\frac{2}{3}\frac{\langle \Sigma \delta v_R \delta v_\phi \rangle}{\langle\Sigma c_S^2\rangle}
\end{equation}
where $\langle \rangle$ denotes an azimuthal average over the entire disc, $\delta v_R=v_R-\langle v_R\rangle$, $\delta v_\phi=v_R-\langle v_\phi\rangle$

\item The gravitational stress $\alpha_G$ which is given by:
\begin{equation}
\alpha_G=\frac{2}{3}\frac{\langle \int_{-\infty}^{\infty}(4\pi G)^{-1}g_Rg_\phi dz\rangle}{\langle\Sigma c_s^2\rangle}
\end{equation}
where $g_R$ ang $g_\phi$ are the radial and azimuthal components of the self-gravitational acceleration. The vertical integration is performed by changing the square of the smoothing parameter $b^2$to   $b^2+\eta^2$ where $\eta$ is such that $z=\eta R$ 
(Baruteau et al. 2010). Following Bae et al. (2014), $\eta$ is varied evenly by $0.01$ from $0$ to $1$.
\end{enumerate}

\section{Evolution in isothermal dics}


\begin{figure}
\centering
\includegraphics[width=\columnwidth]{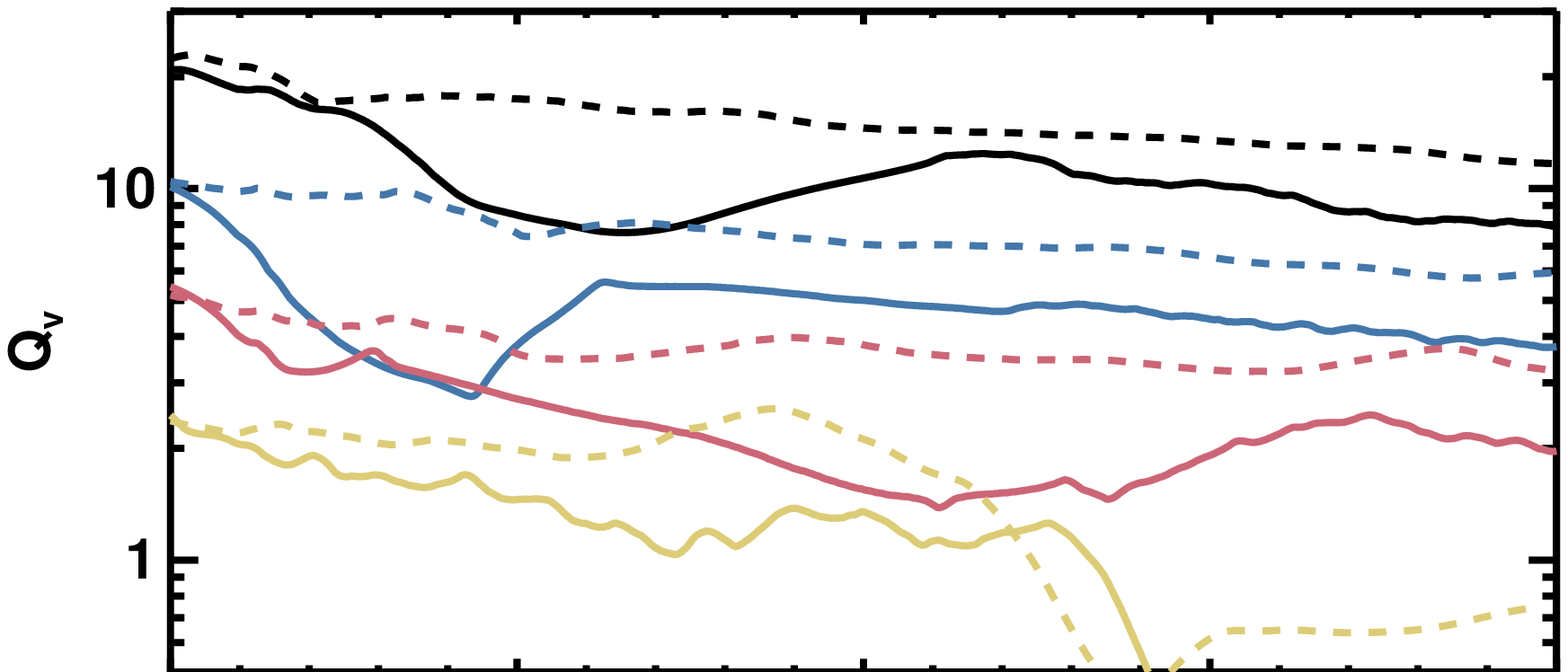}
\includegraphics[width=\columnwidth]{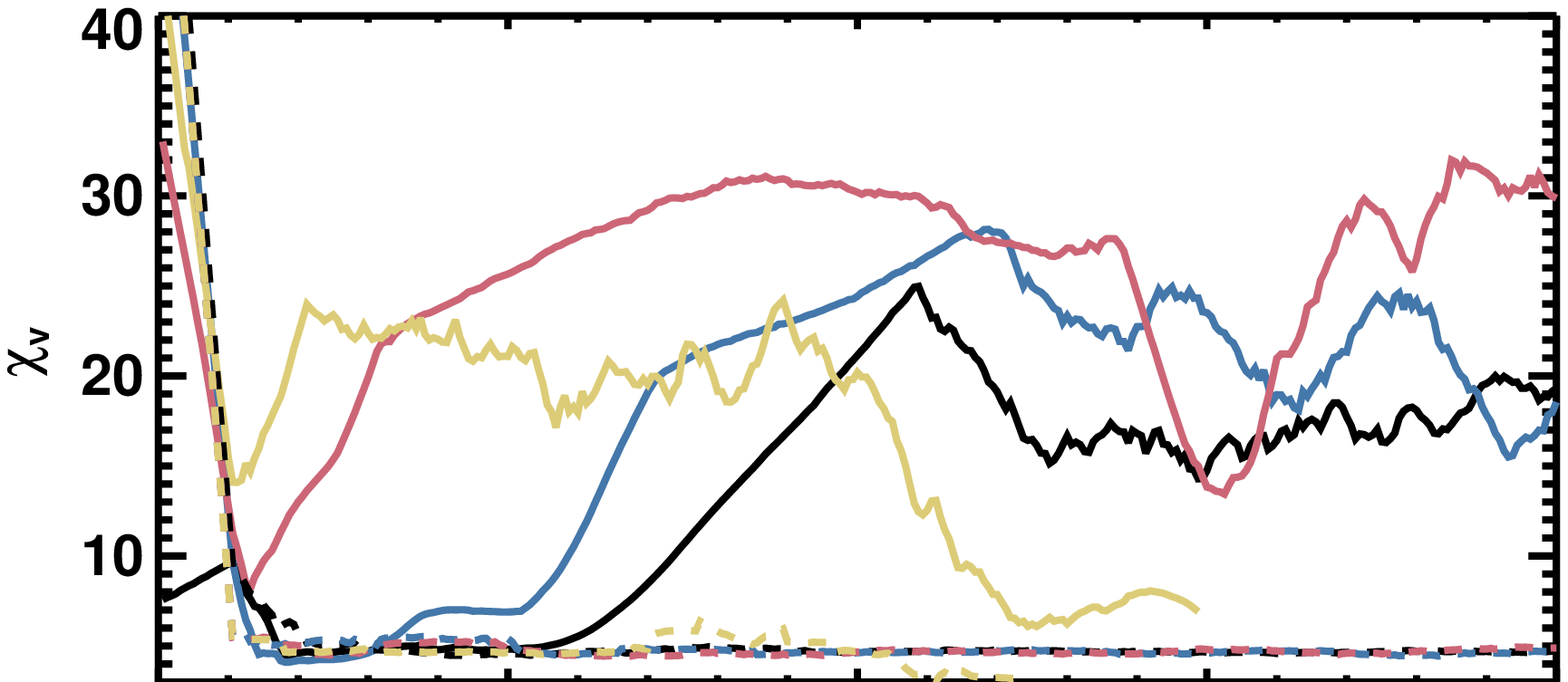}
\includegraphics[width=\columnwidth]{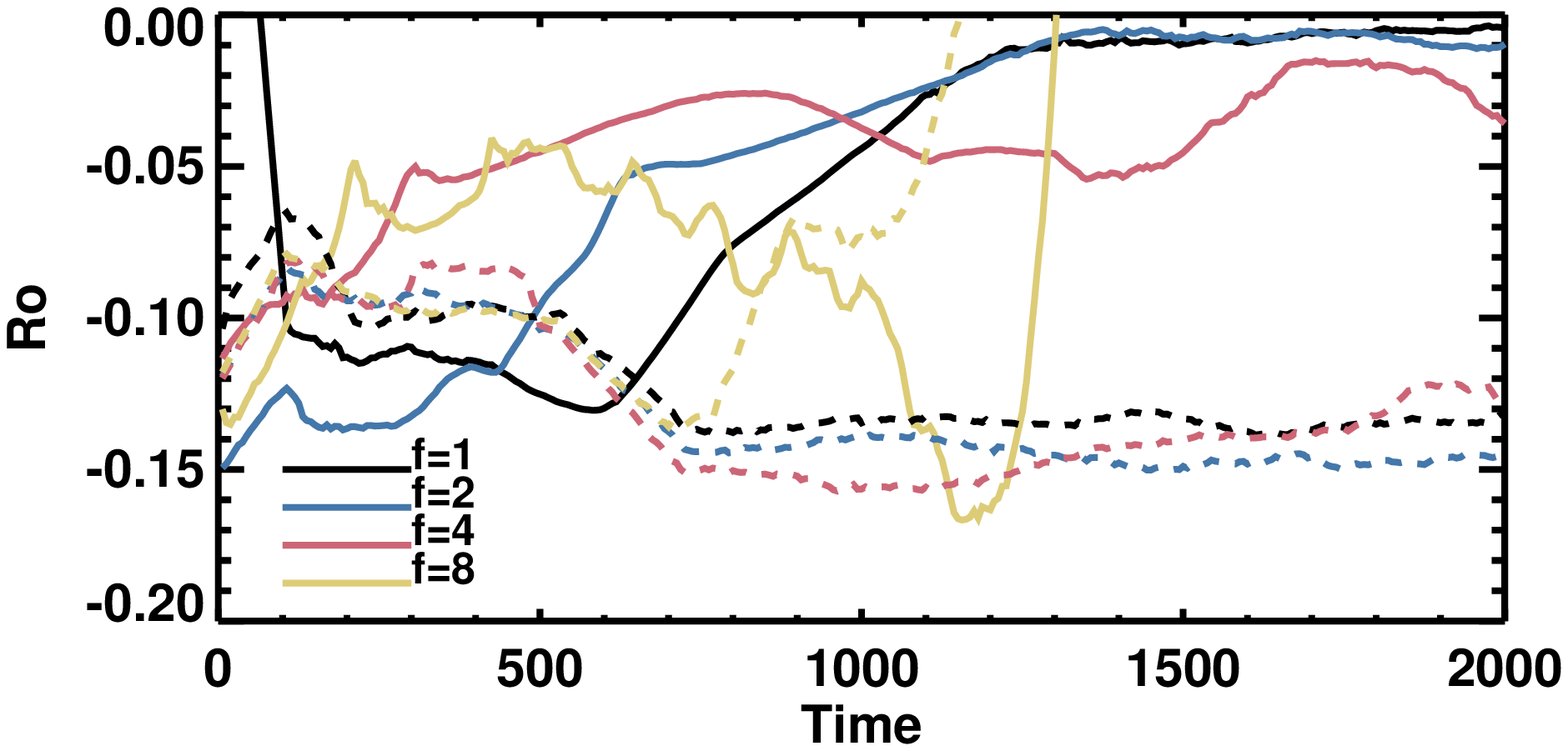}
\caption{{\it Upper panel:} Evolution of the Toomre parameter at vortex centre, averaged over $100$ orbital periods at vortex location, for the isothermal runs with, from top to bottom, $f=1, 2 , 4, 8$. Solid lines correspond to simulations including self-gravity whereas dashed lines correspond to non self-gravitating runs. {\it Middle panel:} same but for the vortex aspect ratio
$\chi_v$. {\it Lower panel:} same but for the  Rossby number calculated at vortex centre.}
\label{fig:qmin}
\end{figure}

\begin{figure}
\centering
\includegraphics[width=\columnwidth]{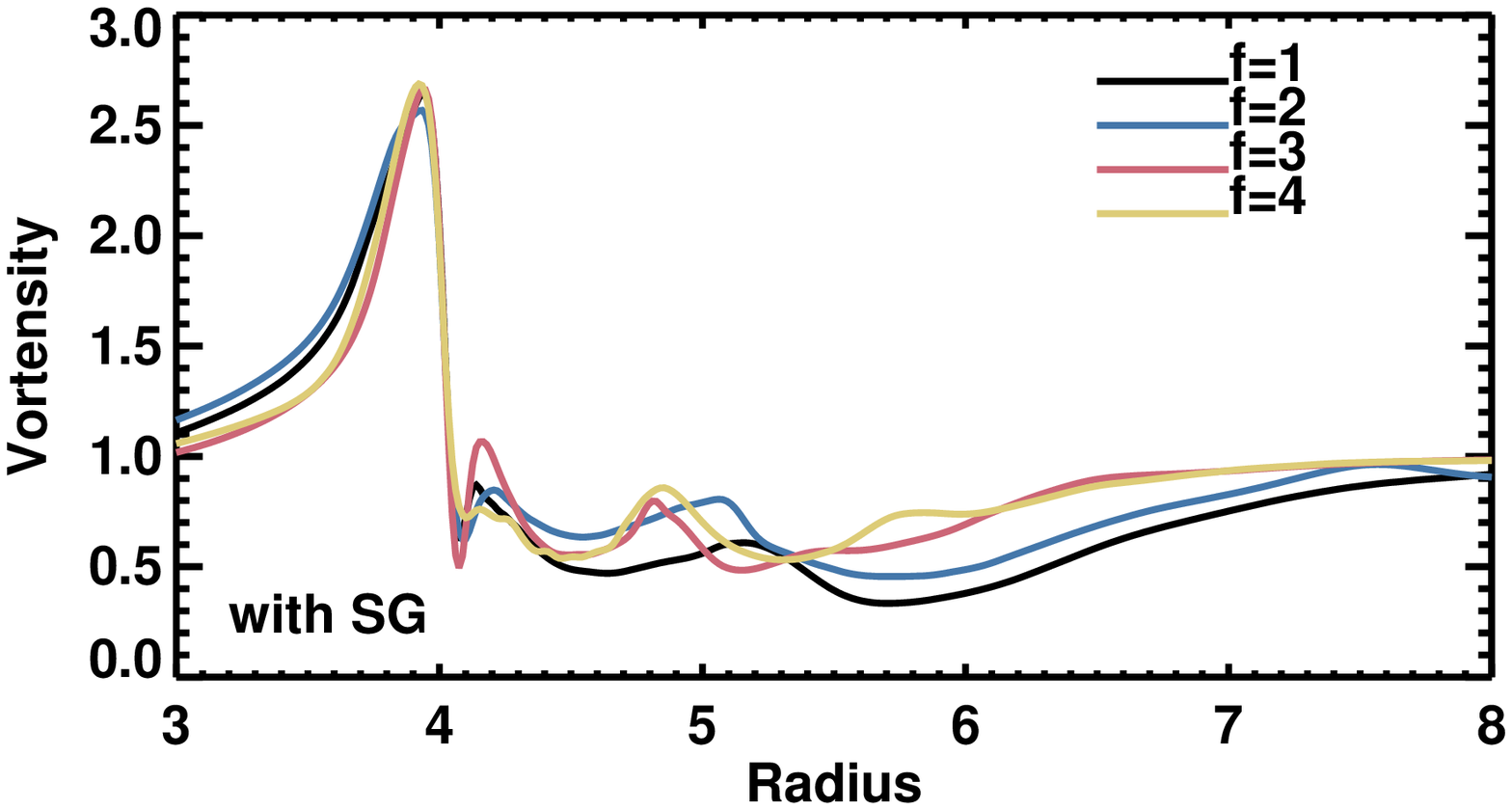}
\includegraphics[width=\columnwidth]{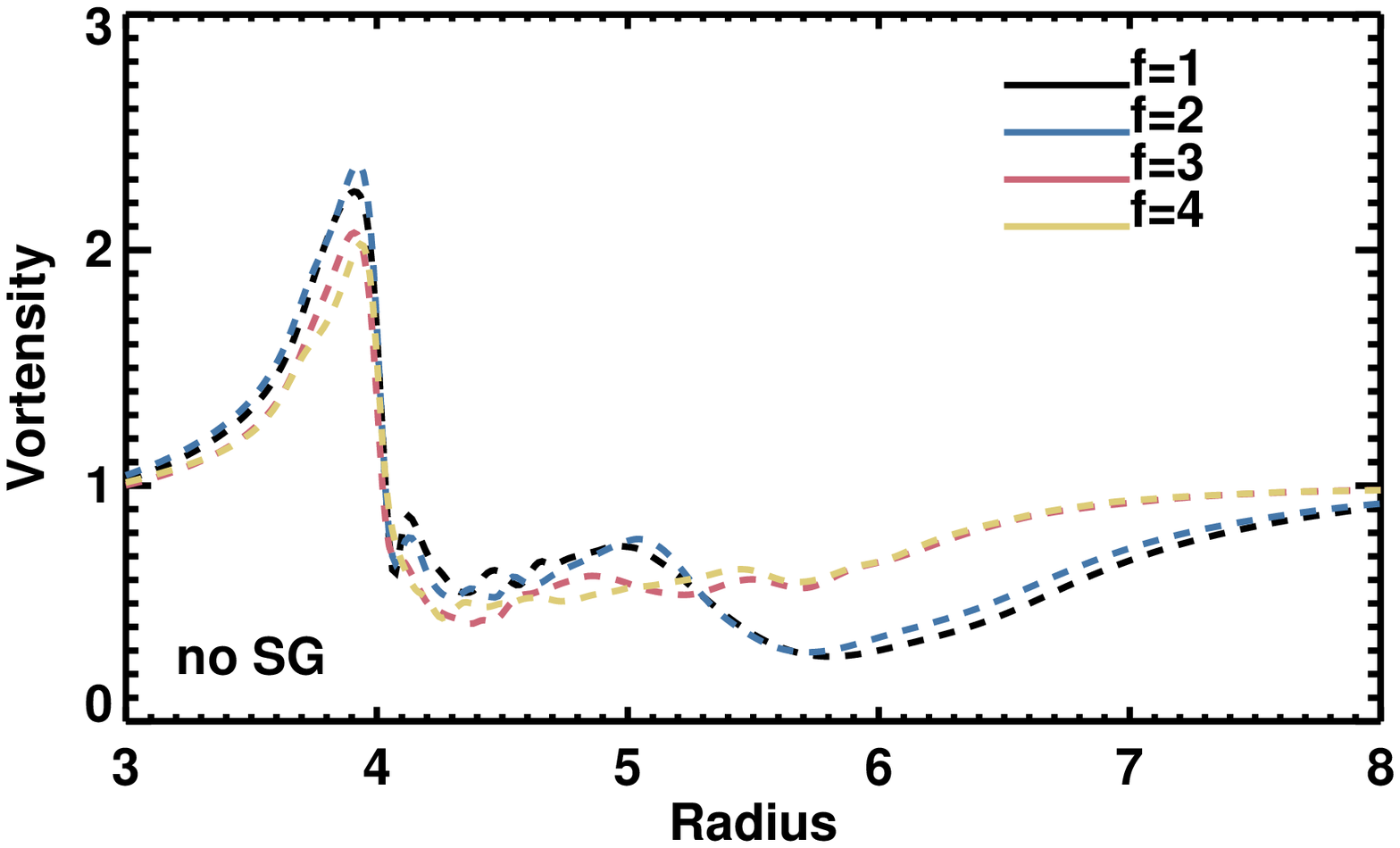}
\caption{{\it Upper panel: }Radial profile of the disc vortensity prior vortex decay occuring, for the isothermal simulations that include self-gravity. {\it Lower panel:} Vortensity profile at the same times but for runs that do not include the effect of self-gravity.}
\label{fig:vort}
\end{figure}

\begin{figure}
\centering
\includegraphics[width=\columnwidth]{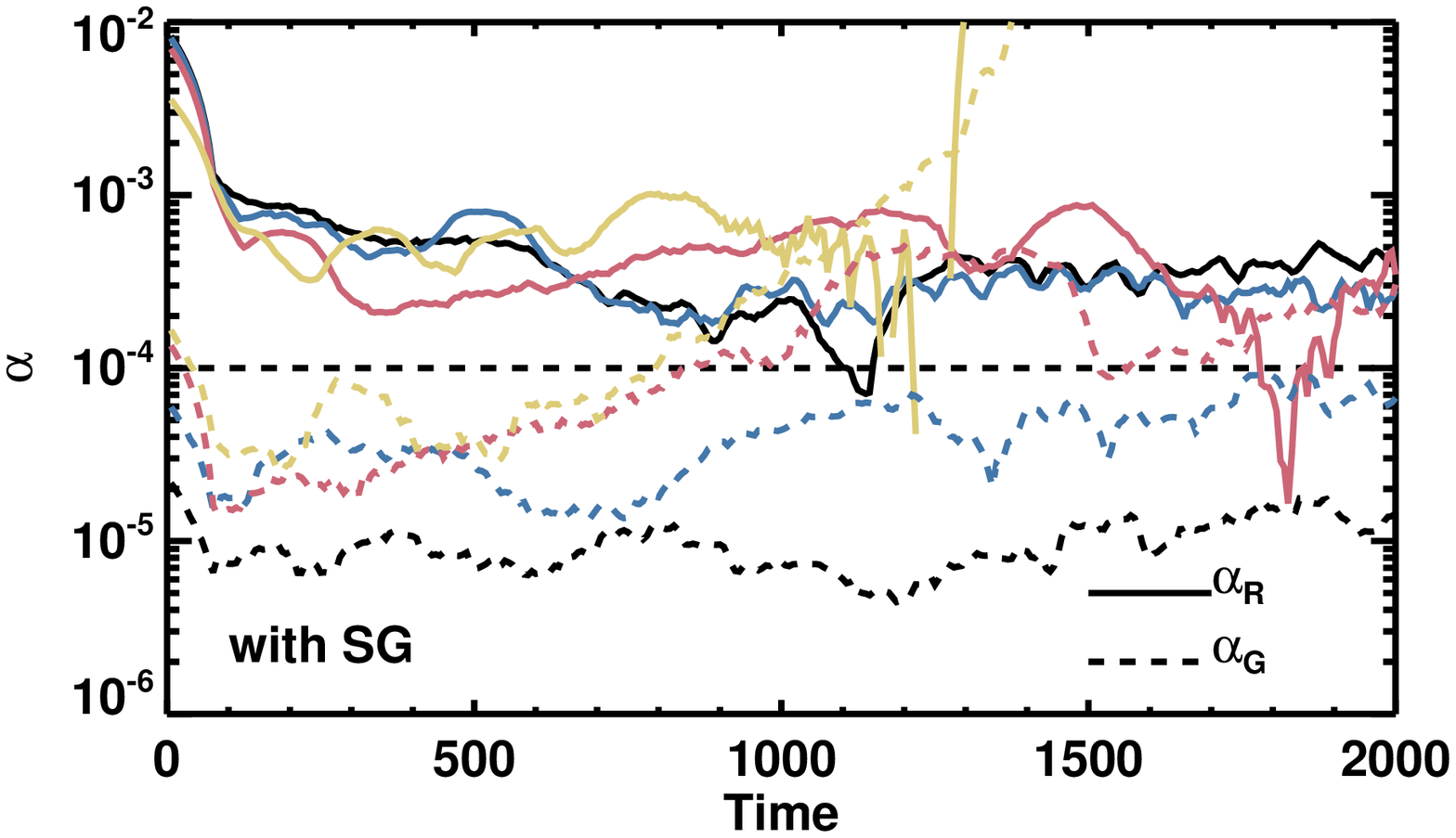}
\includegraphics[width=\columnwidth]{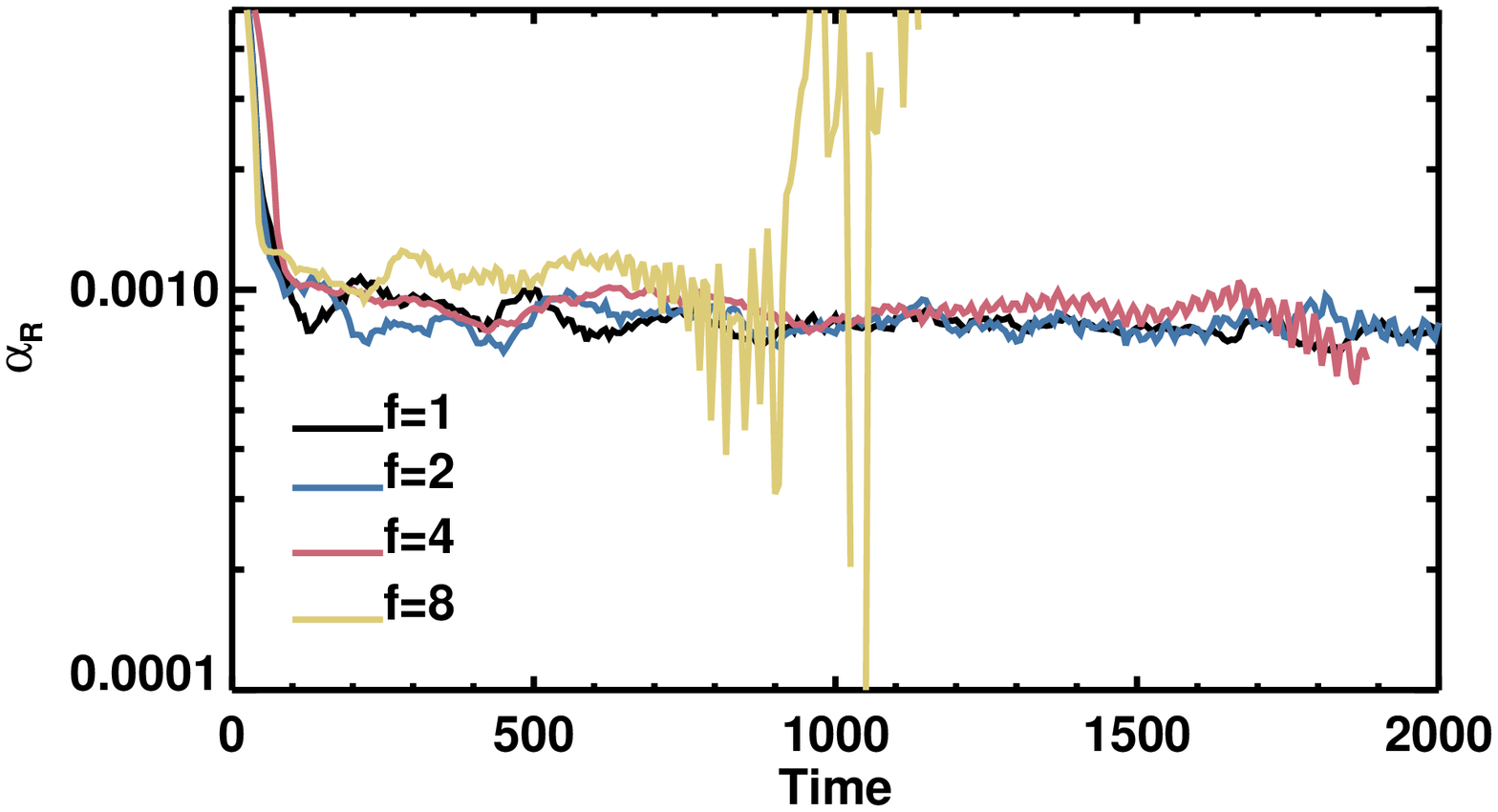}
\caption{{\it Upper panel: } Time evolution of the Reynolds $\alpha_R$ and gravitational stresses $\alpha_G$ for the isothermal simulations that include self-gravity.  The horizontal dashed line corresponds to the value  for the viscous stress parameter employed in the dead-zone. {\it Lower panel:} Time evolution of the Reynolds stress $\alpha_R$ for the runs 
without self-gravity.}
\label{fig:alpha}
\end{figure}

We begin our description of the evolution of vortices in self-gravitating discs by examining the results of the isothermal simulations. In the isothermal limit, we note 
that because the pressure bump acts as a trap for the vortex,  vortex migration is not expected in that case (Paardekooper et al. 2010).\\
  In these runs,  the growth timescale of the  RWI is
 typically $t_{growth}\sim 5$ orbits, is this value is found to be nearly independent on whether self-gravity is taken into account or not.  The most unstable azimuthal wavenumber $m_{max}$ of the RWI, however, is found to be slightly higher in the case where self-gravity is included. For the model with $f=1$, for example, $m_{max}=6$ in the simulation without self-gravity, where $m_{max}=7$ if self-gravity is included. This is exemplified in the first panel of Fig. \ref{fig:run1} which shows, for this model, a snapshot of the disc surface density at $t=5$ orbits.\\
 
 The non-linear evolution of the RWI in simulations with self-gravity,  however, significantly  differs from that in runs where self-gravity is not considered. In the case where self-gravity is discarded, a long-lived 
 vortex with aspect ratio $\chi \sim 4-5$ develops whereas when self-gravity is included,  two different modes of evolution are obtained, depending  on the disc mass. For models with $f=1,2$,  a single vortex that is formed  ultimately decays due to the mechanism described in Regaly \& Vorobyov (2017),  whereas for $f\ge 4$, a vortex with a turbulent core undergoes gravitational contraction. We now describe in more details these two modes of evolution. \\

\begin{figure}
\centering
\includegraphics[width=\columnwidth]{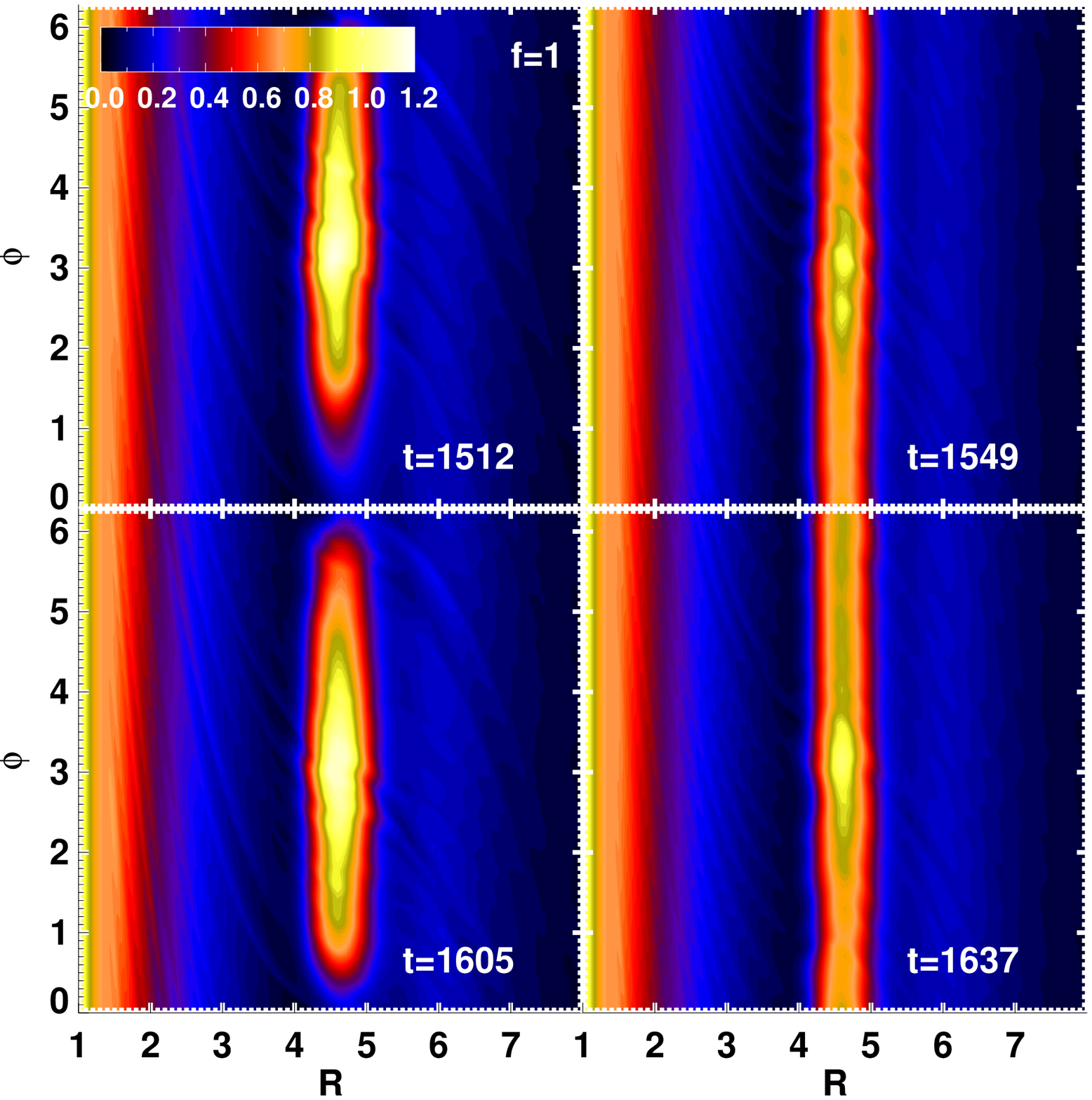}
\caption{Contours of the scaled surface density $\Sigma/\Sigma_{in}$ at different times  for the isothermal, self-gravitating run with $f=1$ and illustrating the cycles of vortex formation--dissipation that are obtained at the end of the simulation.}
\label{fig:run1-qst}
\end{figure}

We use the run with $f=1$ to illustrate the first mode of evolution that we find. For this calculation, contours plots of the surface density at four representative times are displayed in Fig. \ref{fig:run1}. The second panel corresponding to $t=500$ reveals 
that at that time, the initial vortices have merged to form a single vortex with $\chi \sim 5$.  At $t=1000$, however, we see that the vortex aspect ratio has considerably  increased while the azimutal surface density contrast has decreased. 
The surface density structure at $t=1500$ (fourth panel) suggests that the  vortex stretches out until it is completely dissipates in the background flow. This process is further illustrated in Fig. \ref{fig:qmin} where we show  the time evolution of the Toomre parameter at vortex center $Q_v$ (upper panel), vortex aspect ratio $\chi_v$ (middle panel), and Rossby number of the vortex $Ro$ (Lower panel). The continuous decrease in $|Ro|$ in the time period between $600$ and $1200$ orbits confirms that the vortex weakens as vortex stretching occurs. At $t\sim 1200$, $Ro\sim 0$ which indeed indicates that the vortex has been almost entirely suppressed at that time. 
 A new elongated vortex  whose evolution is very similar to that described above eventually emerge at  later times, giving rise thereby to cycles of vortex formation-dissipation. One such cycle  is  illustrated by the sequence of snapshots at four successive times that is presented in  Fig. \ref{fig:run1-qst}. 
We note that similar vortex mode oscillations have also been reported  in previous non self-gravitating simulations of vortices forming at a viscosity transition (Regaly et al. 2012). \\

\begin{figure*}
\centering
\includegraphics[width=\textwidth]{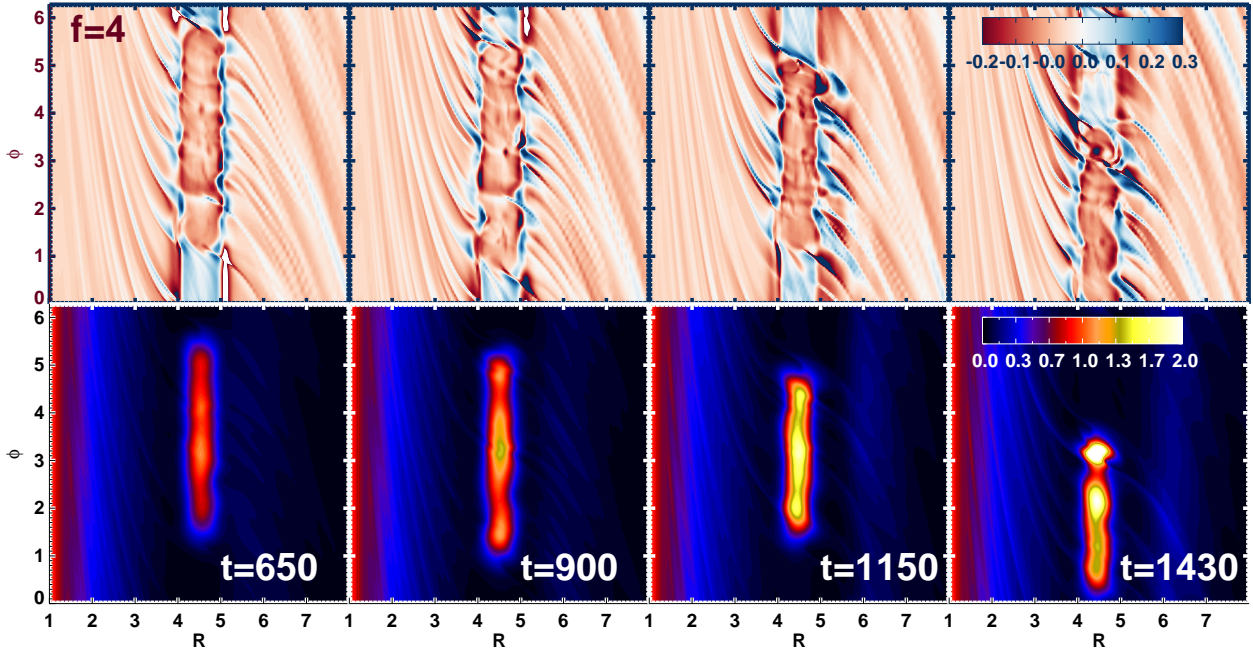}
\caption{{\it Upper panel:} contours of the Rossby number at different times for the isothermal run with $f=4$ and with self-gravity included. {\it Lower panel:} contours of the scaled surface density $\Sigma/(f\Sigma_{in})$ at different times for the same run.}
\label{fig:run3c}
\end{figure*}

\begin{figure}
\centering
\includegraphics[width=\columnwidth]{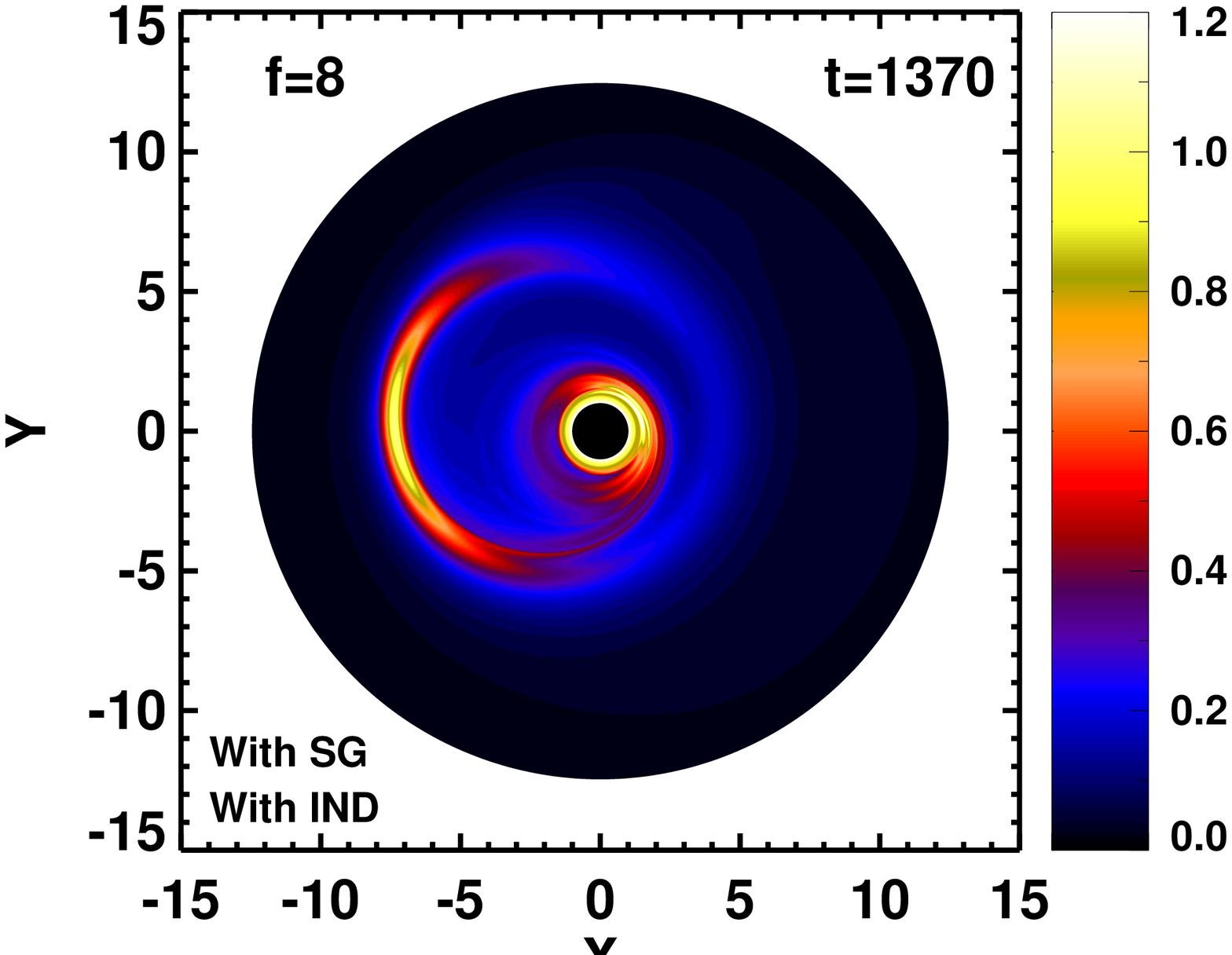}
\includegraphics[width=\columnwidth]{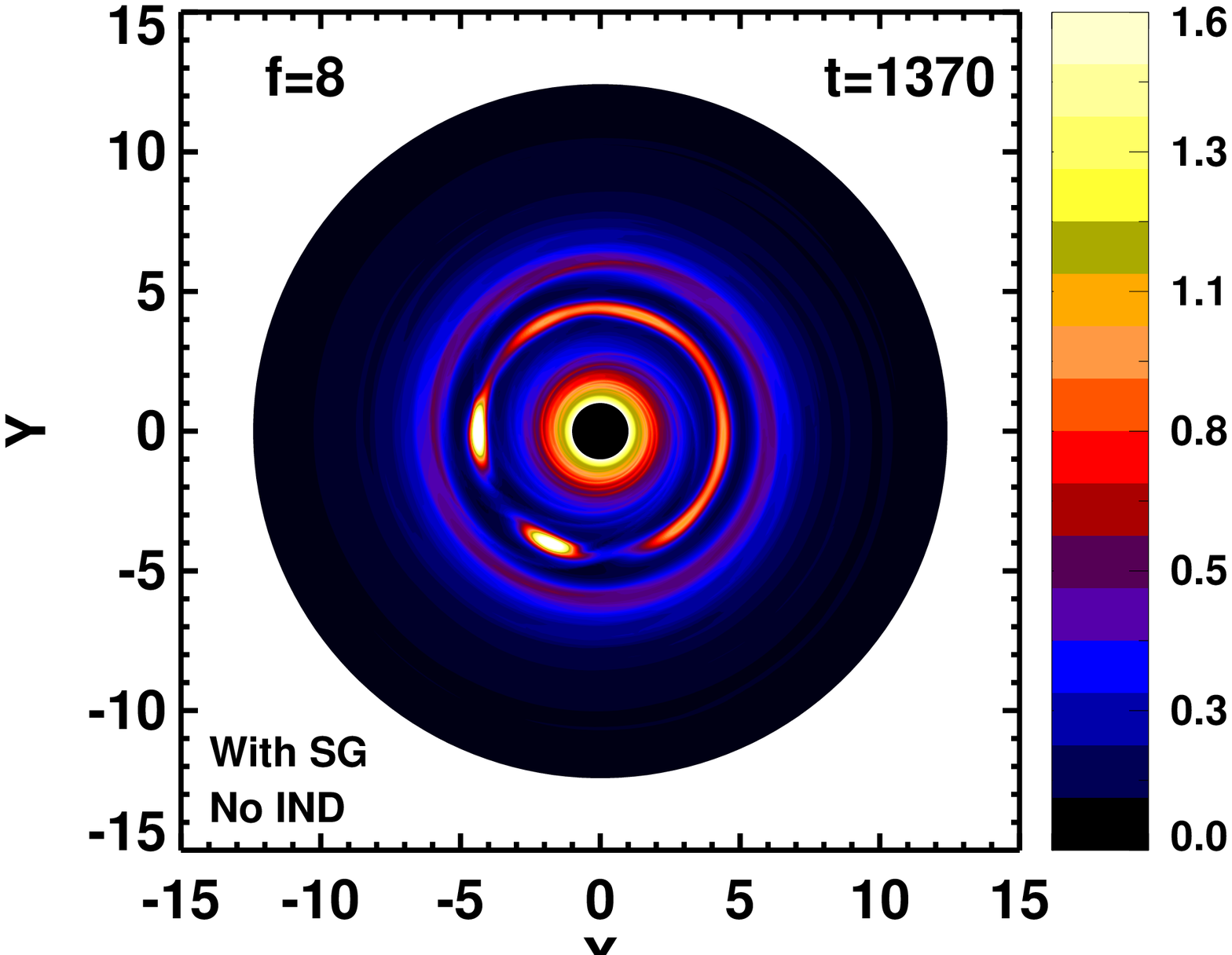}
\caption{{\it Upper panel:} snapshot of the disc surface density at $t=1370$  for the isothermal run with $f=8$ and including self-gravity and the indirect term of the gravitational potential. {\it Lower panel:} same but in the case where the indirect term is discarded.}
\label{fig:ecc2d}
\end{figure}

\begin{figure}
\centering
\includegraphics[width=\columnwidth]{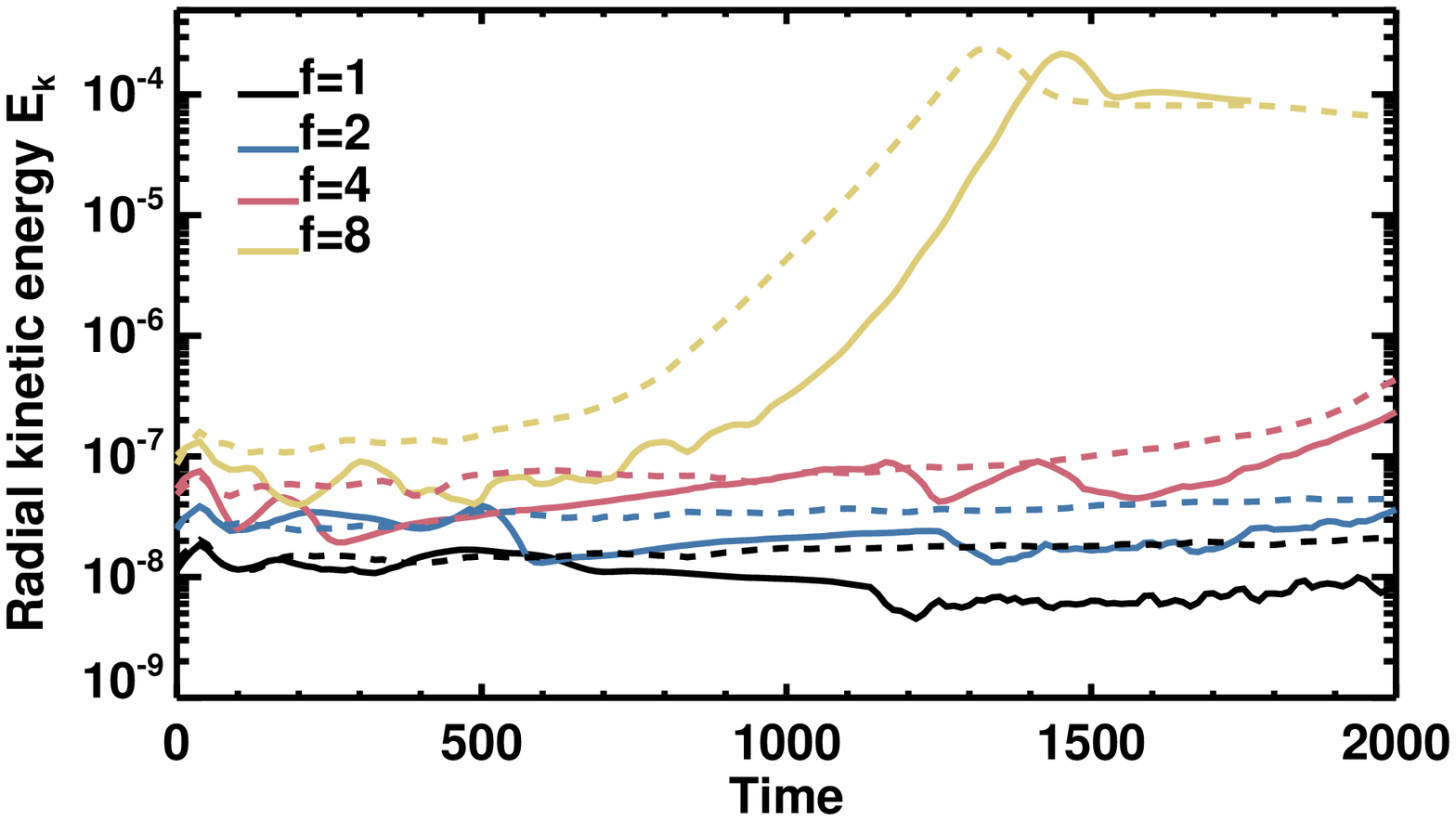}
\caption{Evolution of the total  radial kinetic energy, for the isothermal runs with, from bottom to top,$ f=1, 2 , 4, 8$. Solid lines correspond to simulations including self-gravity whereas dashed lines correspond to non self-gravitating runs.}
\label{fig:ek}
\end{figure}

 The mechanism of vortex decay that is observed in the simulations may arise for two reasons. \\
First, the vortex excites spiral density waves that can turn into shocks once the vortex amplitude becomes high enough. This not only causes the 
vortex to lose energy through shock dissipation, but also can significantly alter the background vortensity  (i.e. the ratio between the vertical component of the vorticity and the disc surface density) profile.   The gap-opening criterion of Crida et al. (2006) predicts that the vortex should be able to carve a gap in the disc provided that ${\cal P}<1$,  where ${\cal P}$ is the gap opening parameter which is given by:
\begin{equation}
{\cal P}=1.1\left(\frac{q_v}{h^3}\right)^{-1/3}+\frac{50\nu}{q_v R_v \Omega_v}
\end{equation}
where $R_v$ is the radial position of the vortex,  $q_v$ is the vortex-to-star mass ratio and $\Omega_v$ the angular velocity at this location. For  the runs with $f=1$ and $f=2$,  we estimate that at the time where the vortex begins to decay $q_v\sim 1.3\times 10^{-4}$ and $q_v\sim 2.8\times 10^{-4}$ respectively. This gives ${\cal P}\sim 0.85$ for the run with $f=2$ whereas ${\cal P}\sim 1.1$ in the case with $f=1$. Therefore, we expect the background vortensity profile to be only weakly altered  in the simulation with $f=1$, while the background vortensity gradient might be significantly smoothed out by the vortex in the run with $f=2$. Compared to the case of a gap carved by a planet, however, we note that  for a vortex the aforementionned gap-opening criterion may also depend strongly on the vortex aspect ratio, since we expect elongated vortices to be much less efficient in modifying the background vortensity profile than nearly circular ones. For the simulations including self-gravity, we show in the upper panel of Fig. \ref{fig:vort} the vortensity profile at the time where the vortex begins to decay,  and which can be different for distint models. For the non self-gravitating case,  the vortensity profiles at the same times are also plotted for comparison  in the lower panel of  Fig. \ref{fig:vort}. Contrary to our own expectation, the non self-gravitating vortex seems to be more efficient in changing the background vortensity profile than the self-gravitating one.  This is confirmed by inspecting the time evolution of the Reynolds stresses, which is presented in Fig. \ref{fig:alpha}. We see that the stresses are slightly higher in the non self-gravitating case, which demonstrates that the spiral waves induced by the non self-gravitating  vortex are more efficient in changing the background vortensity profile.  This is consistent with the non self-gravitating vortex having a smaller aspect ratio, thereby exciting stronger spiral wakes. Despite its ability to modify the background, the  non self-gravitating vortex appears to remain fairly stable over the course of the simulation, such that we can conclude that the mechanism of vortex decay that is observed in the self-gravitating simulations is not related to spiral shocks induced by the vortex.\\
Second, vortex decay can occur due to the self-gravitational torque of the vortex, as described in Regaly \& Vorobyov (2017). These authors showed that the contribution of the self-gravitational torque acting on the leading part of the vortex is negative, whereas the trailing part of the vortex undergoes a positive torque. This, combined with the effect of Keplerian shear, leads to the vortex being continuously stretched out until it completely dissipates in the background flow.  Such a mechanism is estimated to be responsible for vortex decay in discs with masses $M_{disc}/M_\star\gtrsim 0.005$ (Regaly \& Vorobyov 2017),  which is equivalent to $Q\lesssim 50$ at the location of the viscosity transition in the unperturbed disc phase. Here, our runs with $f=1, 2$ in which vortex stretching arises have $Q\sim 30$  and $Q\sim 15$ in the unperturbed disc respectively, which is consistent with the estimation of  Regaly \& Vorobyov (2017).




Simulations with $Q\le 7$  and which correspond to models with $f=4,8$ resulted in a different outcome. As mentionned earlier, the vortex undergoes gravitational collapse in that case. To illustrate how evolution proceeds in that  case, we use the run with $f=4$  and for which the  vortex structure in terms of surface density and Rossby number at 
different times is presented  in Fig. \ref{fig:run3c}.  Similarly to  the run with $f=1$,  merging of initial vortices gives rise to a single vortex whose structure is strongly affected by self-gravity. At $t=650$, 
one can see that the vortex is indeed significantly elongated due to mechanism presented in Regaly \& Vorobyov (2017),  with an aspect ratio  estimated to  $\chi \sim 18$. Contrary to models with $Q\ge 15$, however, the ellipsoidal vortex does not dissipate but rather  strenghens at later times, which is confirmed by inspecting in Fig. \ref{fig:qmin}
the evolution of $Ro$,   which is  continuously decreasing  for $900<t<1100$. This decrease in $Ro$ is accompanied by a 
decrease  in $\chi$, suggesting thereby self-gravitational contraction of the vortex.  As the vortex contracts, the surface density at vortex centre increases, which is unambiguously supported by looking at  the vortex surface density maps at $t=900$ and $t=1150$ in Fig. \ref{fig:run3c}.    Inspection of contours of the Rossby number at $t=1150$ also shows that gravitational 
instabilities can develop within the vortex core.  For simulations including self-gravity, gravitational stresses $\alpha_G$ together with  Reynolds stresses $\alpha_R$ are plotted as a function of time in the upper panel of Fig. \ref{fig:alpha}. We see that  the gravitational stresses continuously increase up to $\alpha_G\sim \alpha_R$, which unambiguously confirms that the vortex core is subject to gravitational 
instabilities.  In appendix  we show that this result is robust regarding both the numerical resolution that is adopted and the amplitude of the density bump at the end of the first step (see Sect. \ref{sec:init})\\
As the vortex  grows, spiral waves launched by the vortex become stronger and stronger.  In Fig. \ref{fig:run3c}, one can  clearly see these strong  wakes extending from either side of the vortex  in the surface density density map corresponding to 
$t=1430$.  These spiral waves contribute to the further gravitational collapse of the vortex by allowing gas accretion onto it. Such a process occurs until a point in time where the vortex becomes strong enough for the spiral waves to turn into shocks, which makes the vortex 
lose energy and decay through shock dissipation (Les \& Lin 2015). The released mass from the vortex subsequently enables the RWI to be re-launched at the location of the pressure bump, which gives rise to the formation of a new vortex whose evolution follows a similar cycle. \\

In disc models with $f=8$, we find that the vortex is massive enough  to make the disc become globally eccentric, as revealed by  looking at contours of the surface density which are plotted at $t=1370$ in {the upper panel of Fig. \ref{fig:ecc2d}}. One possibility to estimate the eccentricity growth within the disc is to compute to radial kinetic energy (Kley \& Dirksen 2006, Teyssandier \& Ogilvie):  
\begin{equation}
E_k=\frac{1}{2}\int_{0}^{2\pi}\Sigma v_R^2 RdRd\theta, 
\end{equation}
whose time evolution is plotted for each run in Fig. \ref{fig:ek}.  Rapid growth of $E_k$ is observed for $f=8$, and this does 
not depend whether or not self-gravity is included. This implies that self-gravity is not the main engine for driving the instability. Instead, it appears that the growth of the disc eccentricity arises  from the effect of the indirect term of the gravitational potential. When the vortex 
is massive enough, we indeed expect the barycentre of the system to be significantly shifted away from the star, which  can cause the development of $m=1$ gravitational 
instabilities (Adams, Ruden \& Shu 1991).   The typical growth rate timescale of such instabilities is expected to not exceed the orbital period as long as $Q\lesssim 3$, which is indeed the case in our runs with $f=8$.   Moreover,  neither growth of the kinetic energy nor growth of the disc eccentricity occured in  simulations that were carried out without the indirect term  included, as can be observed in the lower panel of Fig. \ref{fig:ecc2d} which shows contours of the disc surface density at  $t=1370$ for the run with $f=8$ but in the case where the indirect term for the gravitational potential is discarded. This clearly demonstrates that the indirect term plays a dominant role in triggering the development of the eccentric instability.

\section{Evolution in non-isothermal discs}
\subsection{Models with $\beta$ cooling}

\begin{figure}
\centering
\includegraphics[width=\columnwidth]{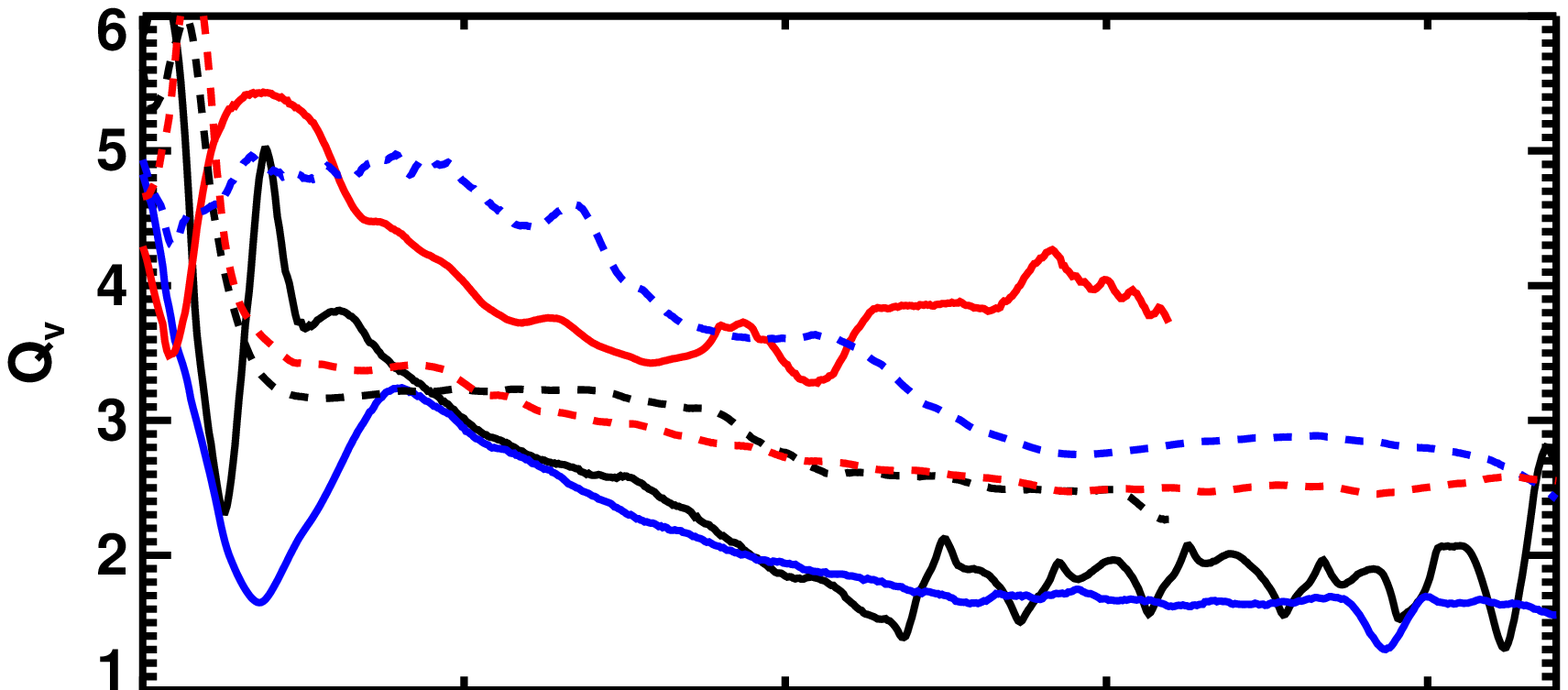}
\includegraphics[width=\columnwidth]{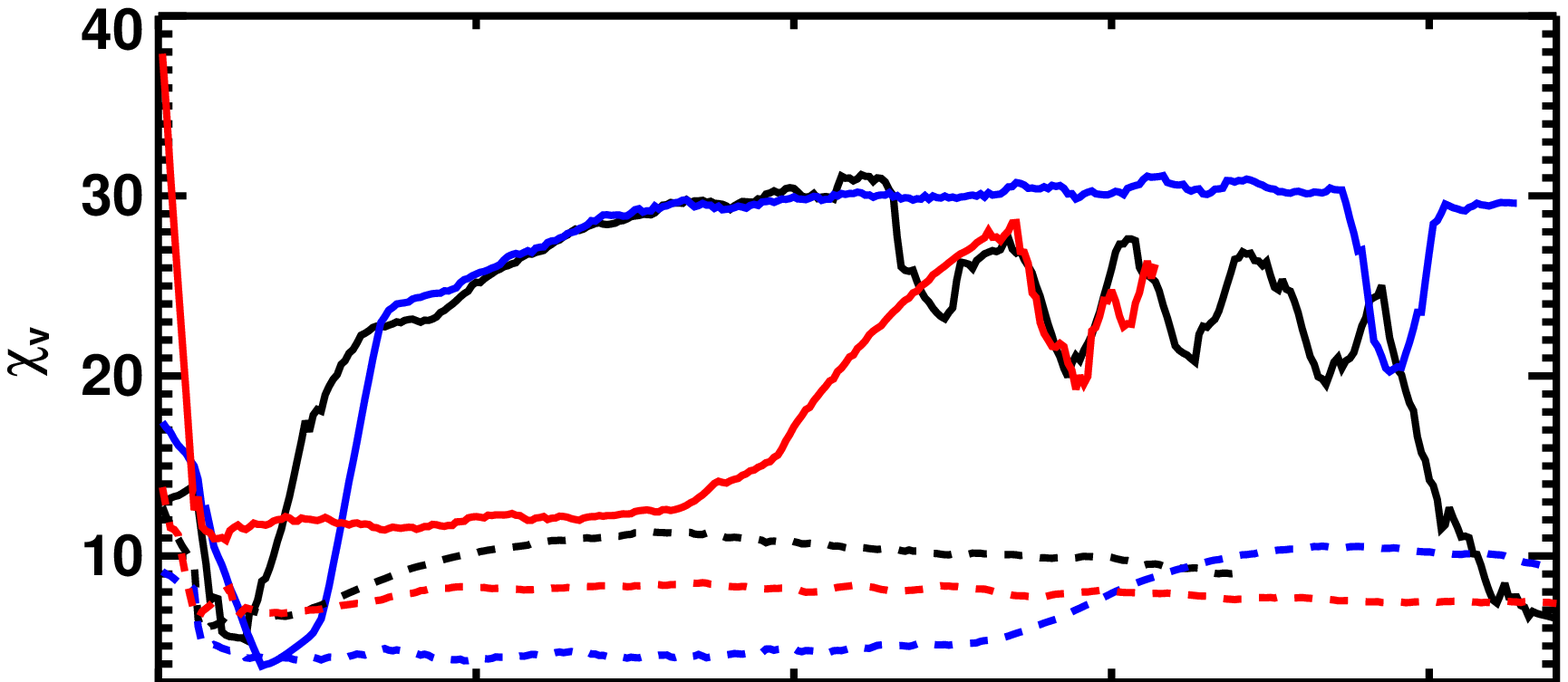}
\includegraphics[width=\columnwidth]{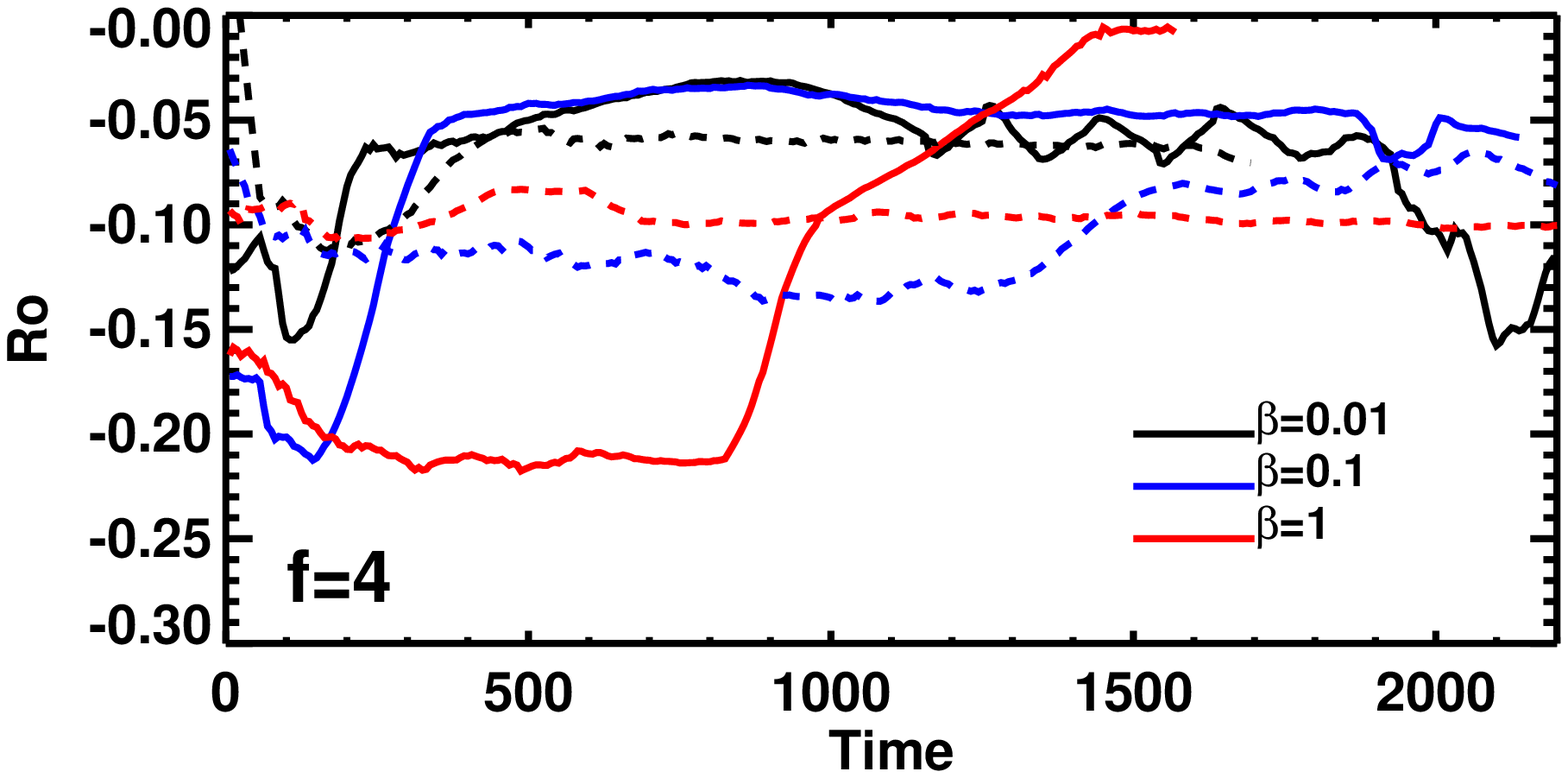}
\caption{ {\it Upper panel:} Evolution of the Toomre parameter at vortex centre, averaged over $100$ orbital periods at vortex location, for the run with $f=4$ and employing the $\beta$ cooling prescription. Solid lines correspond to simulations including self-gravity whereas dashed lines correspond to non self-gravitating runs. {\it Middle panel:} same but for the vortex aspect ratio
$\chi_v$. {\it Lower panel:} same but for the  Rossby number calculated at  vortex centre.}
\label{fig:qminb}
\end{figure}

\begin{figure}
\centering
\includegraphics[width=\columnwidth]{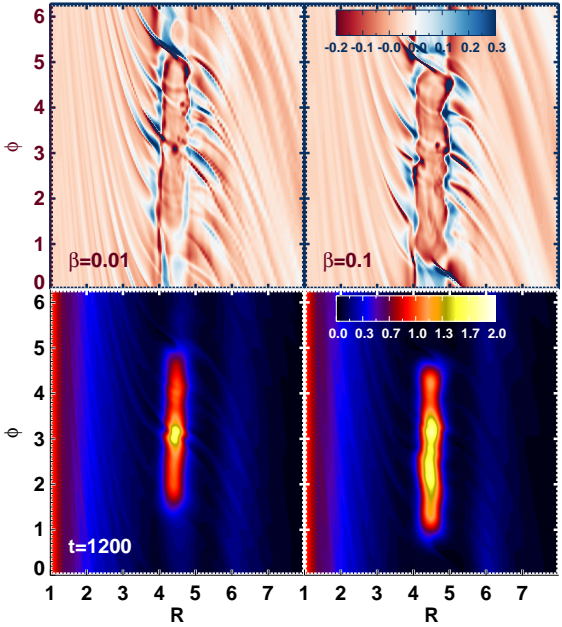}
\caption{{\it Upper panel:} contours of the Rossby number  at $t=1200$ and for the runs employing the $\beta$ cooling 
prescription with $\beta=0.01, 0.1$. {\it Lower panel:} contours of the scaled surface density $\Sigma/(f\Sigma_{in})$.}
\label{fig:run3cb}
\end{figure}

\begin{figure}
\centering
\includegraphics[width=\columnwidth]{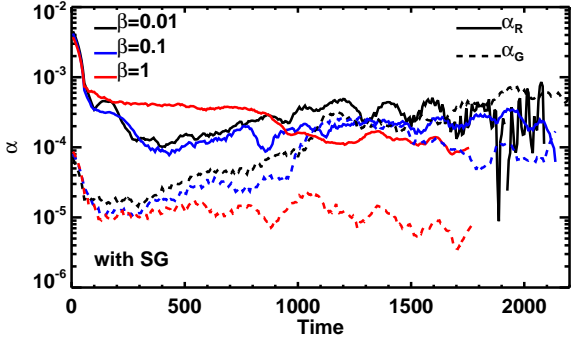}
\caption{Time evolution of the Reynolds $\alpha_R$ and gravitational stresses $\alpha_G$ for runs with $\beta$ 
cooling and  that include self-gravity.}
\label{fig:alphab}
\end{figure}

  In the context of classical gravitational instabilities, it is now widely accepted that the  outcome of such instabilities  strongly depends on the thermodynamical state of the disc.  Assuming a $\beta$ cooling prescription for the cooling timescale 
$\tau_{cool}=\beta \Omega^{-1}$, the disc tends indeed to  achieve a self-regulated state with constant 
$Q$ value for $\beta \gtrsim 10$, whereas lower values for $\beta$  give rise to disc fragmentation. Results from the previous section naturally lead to  the question of how self-gravitating vortices evolve in  discs where the isothermal assumption 
is relaxed. In order to investigate  whether or not  the evolution of such  vortices can be interpreted in the same way as classical gravitational instabilities, we have performed for the case with $f=4$ additional non-isothermal simulations with 
$\beta$ cooling. Simulations that employ a more realistic prescription for cooling will be presented in the next section.  Fig. \ref{fig:qminb} displays, from top to bottom,  the time evolution of the Toomre parameter at vortex 
 centre $Q_v$, vortex aspect ratio $\chi_v$, Rossby $Ro$ number for $\beta=0.01, 0.1, 1$.  In non self-gravitating discs, 
 we can see that there is a tendency for the strength of the vortex to increase with $\beta$, which is consistent with the results of Les \& Lin (2015) who find that the vortex lifetime increases with the cooling timescale. \\
 In self-gravitating discs, the continuous increase in $\chi_v$ and subsequent vortex dissipation that is observed in the 
 run with $\beta=1$ suggests  that the self-gravitating torque is 
at work in that case. We note in passing  that models with $f=1$  (not shown here) also resulted in a similar outcome for $0.01\le\beta\le 1$.   In the case where $\beta \le 0.1$, the results are consistent with the isothermal simulations, with the vortex ultimately collapsing due to the effect of self-gravity. Contrary to the isothermal case, however, we can see that  the Toomre parameter at vortex centre reaches an almost  constant value 
$Q_v\sim 1.5$ prior to the vortex collapsing at $t\sim 2000$ orbits (see upper panel of Fig. \ref{fig:qminb}), which is 
very similar to the value corresponding to a  disc reaching a steady gravito-turbulent state. Inspection of Fig. \ref{fig:run3cb} which shows contours of the Rossby number and surface density at $t=1200$ for both models reveals that gravito-turbulence can indeed develop within the vortex core for $\beta \le 0.1$. The corresponding alpha parameters $\alpha_R$ and $\alpha_G$ asssociated with the Reynolds and gravitational stresses are shown as a function of time in Fig. \ref{fig:alphab}.  During this gravito-turbulent stage, we see that $\alpha_R\sim \alpha_G$, which is consistent with  previous studies of gravito-turbulent discs (Gammie 2001; Baruteau et al. 2011).
The gravito-turbulence operating in the vortex core may be possibly responsible for the vortex collapse observed at later times through the development of secular gravitational instabilities. This occurs because the  anomalous viscosity generated by the  hydrodynamic turbulence  tends to  remove rotational support (Lin \& Kratter 2016). For a 2D, self-gravitating, isothermal disc, the dispersion relation for axisymmetric modes with wavenumbers $k$ is given by (Gammie 1996):
\begin{equation}
s=\frac{\nu k^2(2\pi G |k|-c_s^2k^2)}{\Omega^2+c_s^2k^2-2\pi G \Sigma |k|}
\end{equation}
For $Q\sim 1.5$ and $h=0.05$, $s$ is positive for perturbations whose lenghtscale $L=2\pi/k$ are such that $L\gtrsim H$. The corresponding maximum growth rate $s_{max}$  is given by (Lin \& Kratter 2016):
\begin{equation}
s_{max}=\frac{27\alpha}{16 Q^4} \Omega
\label{eq:smax}
\end{equation} 
For $\alpha=\alpha_R+\alpha_G\sim 10^{-3}$, $s_{max}\sim 3\times 10^{-4} \Omega$ , which corresponds to a 
growth time of $\sim 500$ orbits. This is very  similar to the characteristic time for vortex collapse inferred from the simulations.   However, we caution the reader  that the estimation given by Eq. \ref{eq:smax} results from a linear analysis of secular instabilities that is valid for viscous, Keplerian self-gravitating discs such that it is not clear whether or not it applies to embedded vortices.  Moreover, the averaging procedure may also have some impact on the values obtained for $\alpha_R$ and $\alpha_G$. In fact, we tested the effect of employing an averaging procedure over the vortex surface only and found that stresses are a factor $\sim 1.3$ higher compared to the case where the average is performed over the entire disc. Nevertheless, we can conclude from the above that the slow collapse that is observed is characteristic of secular gravitational instabilities.  \\

\subsection{Consequences on dust trapping}

\begin{figure}
\centering
\includegraphics[width=\columnwidth]{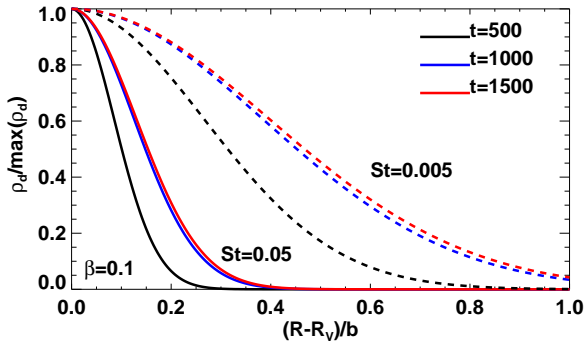}
\caption{Dust density distribution resulting from the model of Lyra \& Lin (2013) and for vortex parameters 
corresponding to the self-gravitating run with $f=4$ and using the $\beta$ cooling prescription with $\beta=0.1$.}
\label{fig:dust}
\end{figure}

The gravito-turbulence arising in the vortex may also have a major impact on dust trapping. To examine this issue in more details, 
we employ the model described in Lyra \& Lin (2013), assuming a GNG model for the vortex.  It has been recently shown that vortices developing at sharp viscosity transitions in self-gravitating discs can be well described by such a model (Regaly \& Vorobyov 2017). According to the model of Lyra \& Lin (2013), the dust distribution within the vortex core is given by:
\begin{equation}
\rho_d(x)=\rho_{d,max}\exp\left(-\frac{x^2b_v^2}{2H_v^2}\right)
\end{equation}
where we set $x=(R-R_v)/b_v$ with $b_v$ the vortex semiminor axis. $H_v$ is the dusty vortex scale height which is given by:
\begin{equation}
H_v=\frac{H}{f(\chi)}\sqrt{\frac{\delta}{St+\delta}}
\end{equation}
 where St is the Stokes number, $f(\chi)$ is a scale function (see Eq. 35 in Lyra \& Lin 2013) and $\delta$ is the dimensionless turbulent diffusion coefficient for which we assume 
$\delta=\alpha_R+\alpha_G$. Focusing on  the model with $\beta=0.1$, we plot  for  $\text{St}=0.005 ,0.05$ the expected dust distribution  at three consecutive times in Fig. \ref{fig:dust}.  Dust grains are initially concentrated close to the vortex centre but as the vortex core becomes gravito-turbulent, increase in the Reynolds and gravitational stresses causes the dust to diffuse away from the vortex centre. We can see that small dust grains with ${\text St}=0.005$ can even be expelled from the vortex due to turbulent diffusion.  Assuming a particle density of $\rho_d=0.8\text{g}\cdot\text{cm}^{-3}$, this would correspond to  dust particles with typical radius $a_d\sim 1.5$ cm.

\subsection{Radiative disc models}

\begin{figure}
\centering
\includegraphics[width=\columnwidth]{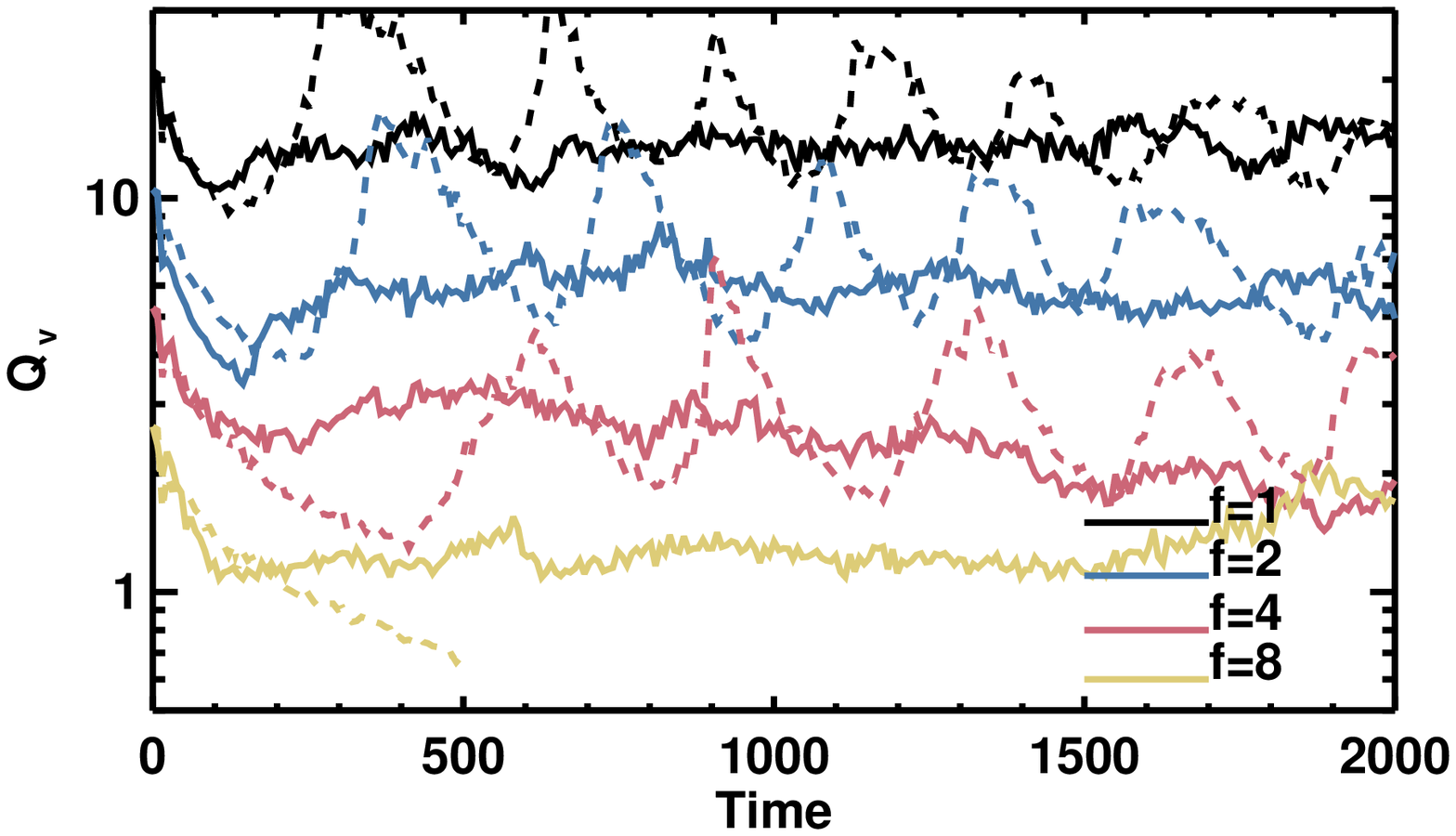}
\includegraphics[width=\columnwidth]{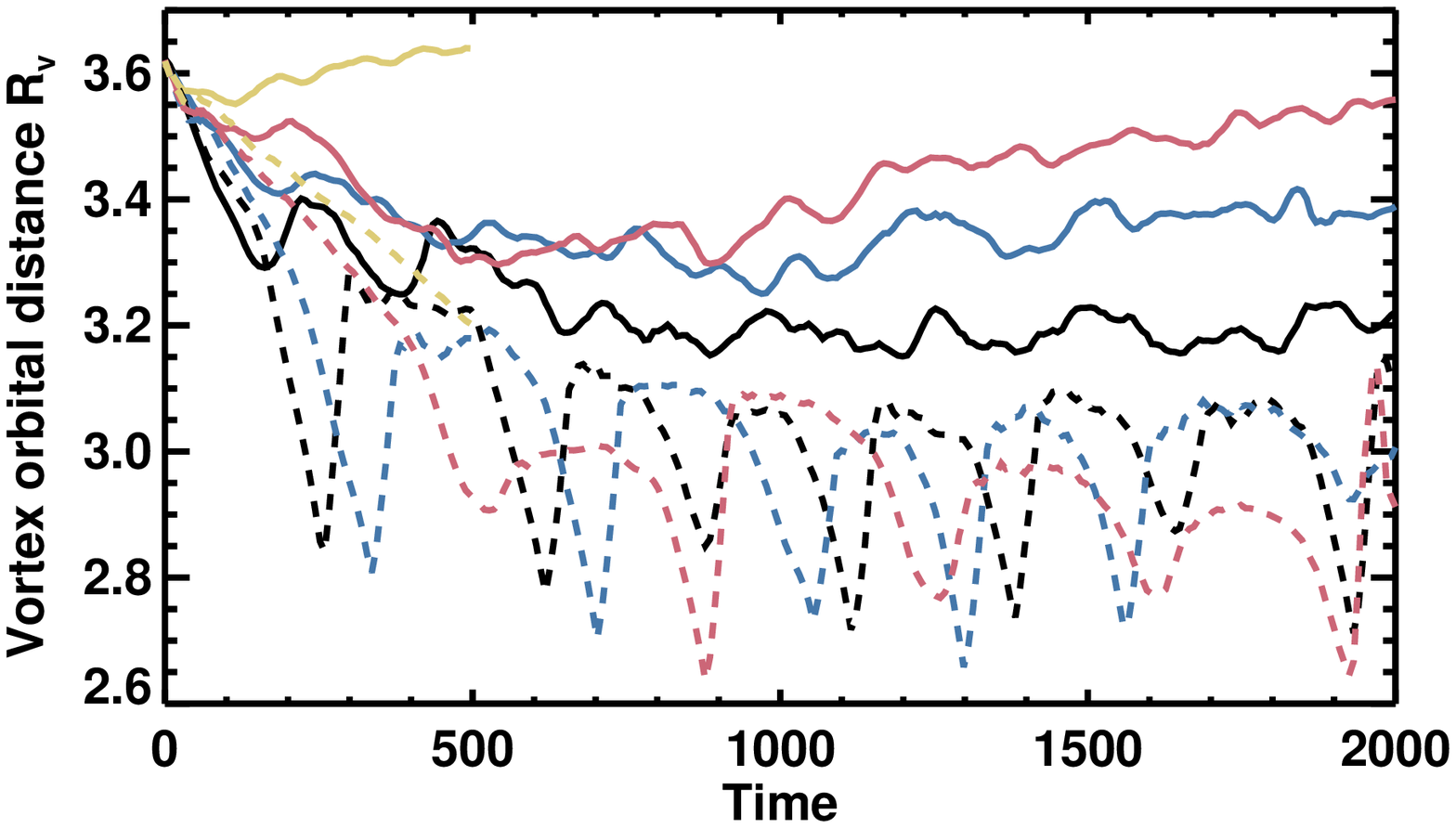}
\caption{{\it Upper panel:} evolution of the Toomre parameter at vortex centre, for the radiative runs with, from top to bottom, $f=1, 2 , 4, 8$. Solid lines correspond to simulations including self-gravity whereas dashed lines correspond to non self-gravitating runs. {\it Lower panel:} time evolution of the vortex radial position for the same runs.}
\label{fig:dvortex}
\end{figure}

 \begin{figure*}
\centering
\includegraphics[width=\textwidth]{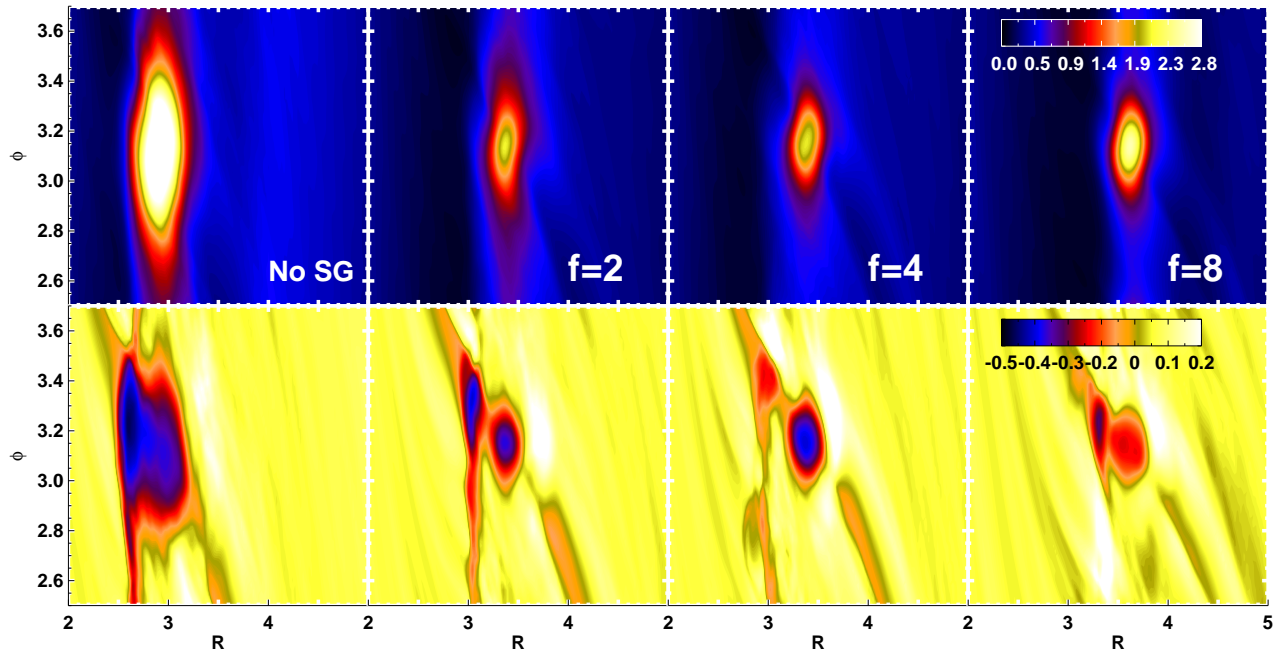}
\caption{{\it Upper panel:} contours of the scaled surface density $\Sigma/(f\Sigma_{in})$ at quasi-steady state for the radiative runs. {\it Lower panel:} Map of the  temperature perturbation  (relative to the azimuthally-averaged temperature) for the same runs. }
\label{fig:2drad}
\end{figure*}

\begin{figure}
\centering
\includegraphics[width=\columnwidth]{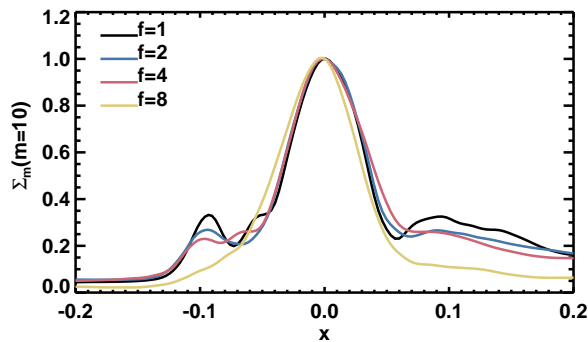}
\caption{$m=10$ Fourier component of the surface density as a function of $x=(R-R_v)/R_v$ for the radiative runs with self-gravity.}
\label{fig:Fourier}
\end{figure}

We now consider how the evolution of vortices proceeds in a radiative disc, employing a more realistic treatment for the cooling term. For a viscosity function that 
depends explicitly on temperature (see Eq. \ref{eq:alpha}) and gas cooling function given by Eq. \ref{eq:cooling}, Faure et al. 
(2015) have reported cycles of  formation, migration and disruption of vortices forming at a pressure maximum. Provided that the vortex does not significantly alter the background vorticity profile, migration of vortices is a priori  not  expected in that case (Paardekooper et al. 2010). Faure et al. (2015) suggest that vortex migration is associated  with a baroclinic 
term in the vorticity equation, possibly arising from a strong  azimuthal temperature gradient. As the cold vortex migrates and penetrates within the much warmer  active region, it becomes progressively eroded through diffusion effects,  until the released mass gives rise to  a new vortex which follows the same evolution. \\
 For our non-isothermal models with $f=1,..,8$,  the upper panel of Fig. \ref{fig:dvortex} shows  the Toomre parameter at vortex centre   as a function of time. Similarly to the isothermal case, the run with 
$f=8$ and in which self-gravity is not included resulted in the growth of the disc eccentricity due to the development of a strong $m=1$ gravitational  instability whereas for the other non self-gravitating  models with $f\le 4$,  $Q_v$ exhibits an 
oscillating behaviour due to the aforementionned vortex cycles.  In the case where self-gravity is included, however, $Q_v$ rather reaches a constant value which suggests that  self-gravity can  stabilize the vortex against baroclinic effects.  Here, the cooling timescale is estimated to be $\tau_{cool}\sim 1/b T^3\sim 1/\alpha \Omega\sim 16 T_{orb}$ which would 
correspond to $\beta \sim 100$ using the standard $\beta$ parametrization. The results of these radiative runs  therefore suggest that the vortex's fate is strongly influenced by the details of thermodynamics,  since a non-isothermal model with $\beta$ cooling would give rise to vortex decay for $\beta=100$. We show  in Fig. \ref{fig:2drad} contour plots of the 
normalized surface density (upper row) and   temperature perturbation relative to the azimuthally-averaged temperature (lower row) for various runs namely: i)  for the case with $f=1$ and without self-gravity included, prior inward migration of the vortex proceeding and ii) for the self-gravitating discs with $f=2,4,8$, once the vortex has reached a quasi-equilibrium structure. Comparing the density maps for  the two  self-gravitating models with $f=4$ and $f=8$, it is clear that the vortex becomes denser in more massive 
discs. As the disc mass increases,  however, the temperature inside the vortex also increases, as revealed by inspecting the temperature contours for the two models, which consequently leads to the vortex reaching a constant $Q_v$ value. \\

 The lower panel of Fig. \ref{fig:dvortex} shows for each model the radial position of the vortex $R_v$ as a function of time.  It is immediately evident that the  vortex cycles are  suppressed when self-gravity is taken into account. Consistently with previous simulations of self-gravitating vortices (Zhu \& Baruteau 2016), the vortex is observed to even migrate outward in self-gravitating runs with $f\ge 4$. From the maps of the relative temperature perturbation in Fig. \ref{fig:2drad},  we can envision two possible mechanisms that are responsible for the vortex outward migration. First, one can see that the vortex's inner wave tends to be eroded in runs with $f\le 2$, mainly at the location which corresponds to the transition  $T\sim T_{MRI}$ . In models with $f\ge 4$, 
 however, the inner wave is much less sheared out, probably because the vortex core is warmer in that case. We note that the inward migration of the non self-gravitating vortex may be partly due to this process, since it tends to increase the 
 asymmetry between the inner and outer waves of the vortex. \\
  Changes in the 
 structure of the outer wake with disc mass can also be identified from these temperature maps, but become more appearant  
 when looking at the  Fourier components of the surface density. For the various runs with self-gravity included,  the $m=10$ component of the surface density   $\Sigma_m$ is plotted as a  function of  $x=(R-R_v)/R_v$  in Fig. \ref{fig:Fourier}.  We choose such a value for $m$ because in the context of planetary migration, the outer Lindblad torque peaks approximately at $m\sim 10$ for a disc with $h=0.05$ (e.g. Ward 1997). As the disc mass
 increases, there is a clear trend for the disc response in the outer disc to decrease and for the vortex to become more asymmetric. This tends to reinforce the effect of the vortex's inner wave, and to favor consequently the outward migration of the vortex.

\section{Discussion and conclusion}

In this paper, we have presented the results of 2D hydrodynamical simulations that examine the role of self-gravity on the long-term evolution of vortices . These vortices form through the development of the RWI at a pressure bump, which  is assumed to be located at the inner edge of a dead-zone, where there is a sharp transition in the viscosity parameter $\alpha$. We focused on the case of massive protoplanetary discs where the Toomre parameter at the location of the pressure bump is initially $Q\le 30$,  and considered both isothermal and non-isothermal equations of state. \\
For isothermal discs, we find  three different modes of evolution depending on the initial value for $Q$. \\
i) For $Q\ge 15$, 
self-gravity makes a large-scale vortex decay due to the self-gravitational torque of the vortex, as predicted by 
Regaly \& Vorobyov (2017).  Once the vortex has completely dissipated in the background flow,  a new vortex emerges at the location of the pressure bump and which follows the same evolution, giving rise  to cycles of vortex formation-dissipation. \\
ii)  Isothermal models with $3\lesssim Q \lesssim 7$ result in the formation of an elongated vortex with aspect ratio 
$\chi \sim 20-30$ and whose core is found to be turbulent due to the development of gravitational instabilities. At later times,  however,  the vortex strenghens due to gravitational collapse but is ulimately found to   decay  once the spiral waves launched by the vortex become too strong. Eventually,  a new vortex can  emerge at the location of the pressure bump and which follows a similar evolution. \\
iii) More massive 
discs with $Q\lesssim 3$  become globally eccentric  due to the growth of a $m=1$ mode in the disc.  In that case, 
it appears that destabilization of the system is not caused by the effect of self-gravity but rather occurs because the vortex is massive enough to significantly shift the barycentre of the system away from the central star.\\
Vortex decay is also observed in   $\beta$ cooling discs with $\beta \ge 1$ due to combined effect of self-gravity and Keplerian shear. In models with $\beta \le 0.1$,  gravito-turbulence can operate  in the vortex core  in models with $3\lesssim Q \lesssim 7$,  with the core maintaining a constant $Q_v$ value. Similarly to the isothermal case, the vortex collapses at later times, possibly because anomalous viscosity arising from gravito-turbulence within the disc tends to remove rotational support. \\
Regarding radiative disc models where the viscosity depends on a switch to disc temperature, non self-gravitating simulations resulted in cycles of vortex formation-migration-disruption,  in good agreement with previous work  (Faure et al. 2015).  Including self-gravity however,  causes the vortex cycles to be completely suppressed. This was found to occur because the vortex tends to be warmer and more asymmetric in self-gravitating discs.  In that case, a quasi-steady state is reached for which the vortex lasts for ${\cal O}(10^3)$ orbits and has aspect ratio of $\chi_v \sim  3-4$. \\
Our results suggest  that self-gravity makes difficult forming long-lived vortices at a pressure bump, both in isothermal disc and non-isothermal discs that employ the $\beta$ cooling prescription. In a radiative disc model where is implemented a better treatment of thermodynamics, however, self-gravity can give rise to a stable vortex structure with aspect ratio $\chi_v  \sim 3-4$.  Strong vortices with aspect ratio $\chi <4$ are known to be destroyed by the elliptical instability (Lesur \& Papaloizou 2009; Richard et al. 2013 ).  Moreover, we note that in this work vortices are formed on a timescale corresponding to $\sim 10$ orbits, which is comparable to the elliptical instability growth timescale in 3D discs.  Therefore, it is not clear whether or not a  vortex formed at a pressure bump might be able to survive to  the elliptical instability , and  3D simulations  are clearly required to investigate this issue in more details. 3D simulations including self-gravity and  that examine the possible role of the elliptical  instability on the evolution of vortices, together with the effect of self-gravity on the elliptical instability will be presented in a future study (Lin \& Pierens, in prep.).

\appendix
\label{sec:app}
\section*{Appendix: Numerical issues}

\begin{figure}
\centering
\includegraphics[width=\columnwidth]{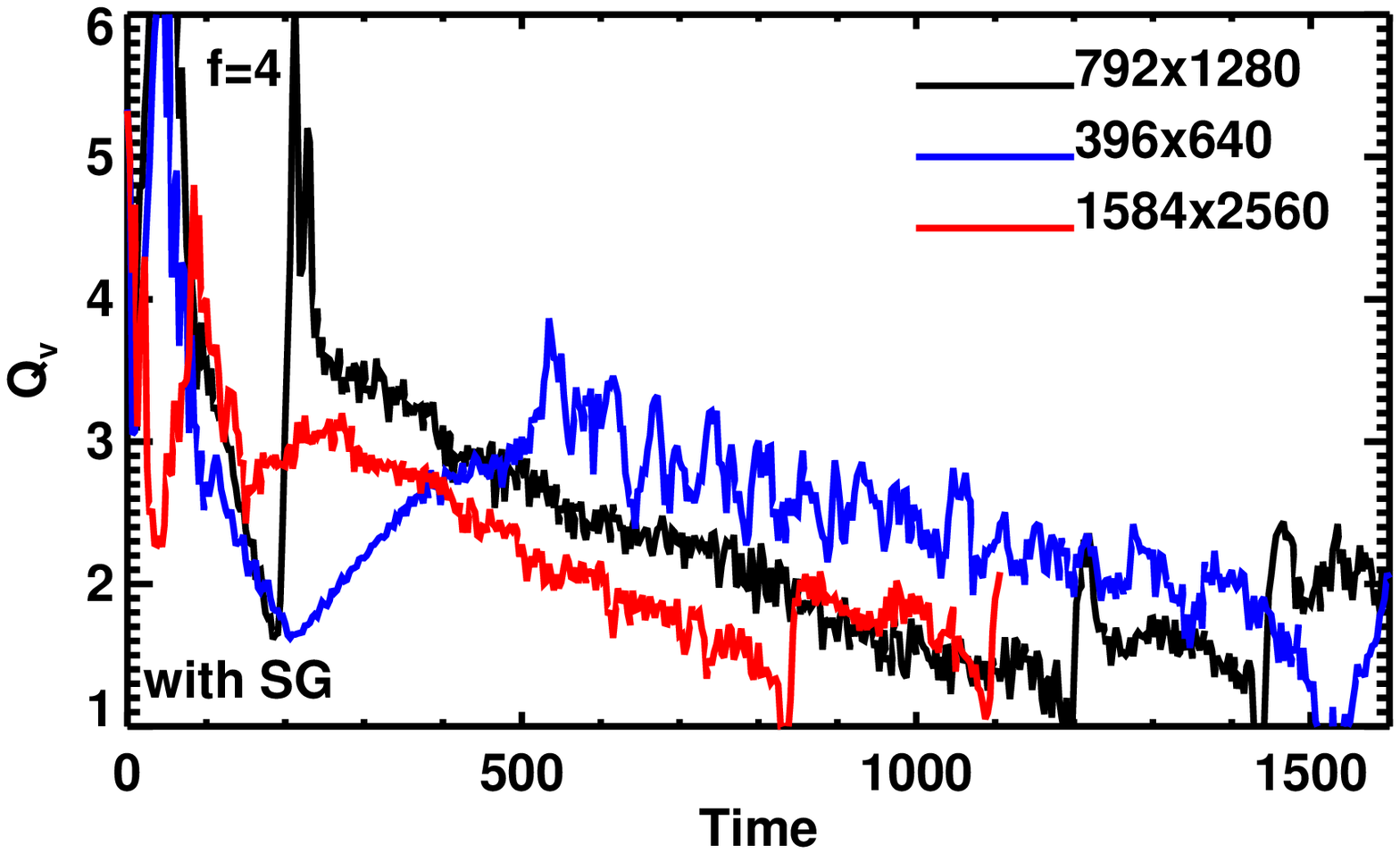}
\includegraphics[width=\columnwidth]{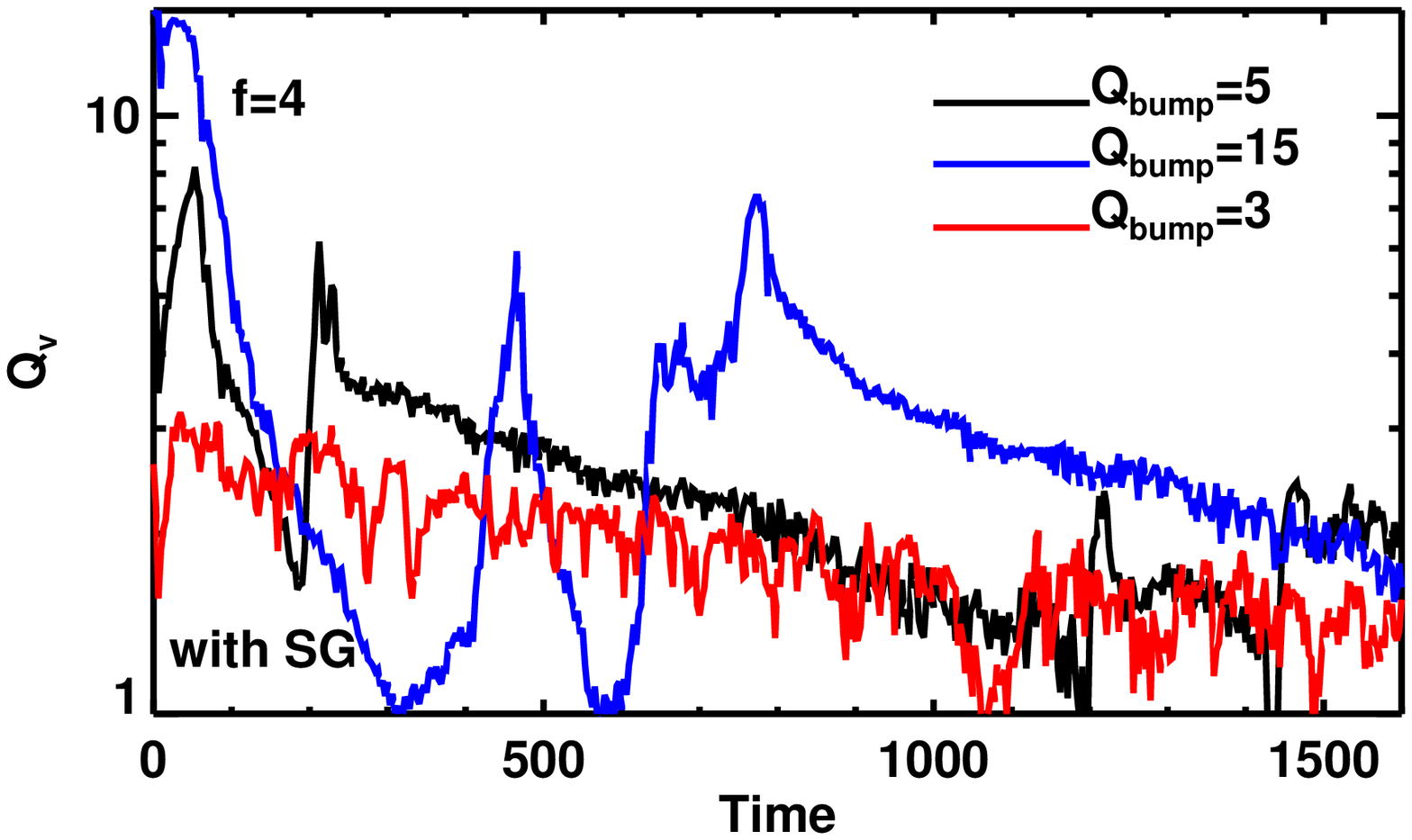}
\caption{{\it Upper panel} Time evolution of the Toomre parameter at vortex centre $Q_v$ for the self-gravitating model with 
$f=4$, and for different  grid resolutions. {\it Lower panel:} $Q_v$ as a function of time for the same run with our nominal grid resolution, but for different initial amplitudes for the density bump.  }
\label{fig:conv}

To assess the impact of numerics on the results presented in this paper, we performed a suite of additional simulations for various resolutions and different amplitudes for the initial density bump. The upper panel of Fig. \ref{fig:conv} shows, for the self-gravitating model with $f=4$, the time evolution of the Toomre parameter at vortex centre $Q_v$ for runs with our nominal resolution ($792\times 1280$), increased grid resolution ($1584x2560$), and lower grid resolution ($396\times 640$). These three simulations exhibit different relaxation phases, but all ultimately resulted in vortex growth, with a very similar corresponding growth timescale. \\
We plot in the lower panel of  Fig. \ref{fig:conv} $Q_v$ as a function of time for the same model and nominal grid resolution, but  for runs which differ by the  amplitude of the density bump at the end of the first step (see Sect. \ref{sec:init}), and therefore by the initial value for the Toomre parameter  $Q_{bump}$ at the location of the pressure bump. The case $Q_{bump}=15$ corresponds to a situation where the initial amplitude of the density bump is zero, such that the vortex forms as soon as the density bump is strong enough for the RWI criterion to be satisfied.  We see that the duration of the relaxation phase is longer in that case, but vortex collapse ultimately occurs, consistently with what found in our reference model  and in the model with $Q_{bump}=3$.

\end{figure}
\section*{Acknowledgments}
We thank Z. Regaly for useful discussions. Computer time for this study was provided by the computing facilities MCIA (M\'esocentre de Calcul Intensif Aquitain) of the Universite de Bordeaux and by HPC resources of Cines under the allocation A0010406957 made by GENCI (Grand Equipement National de Calcul Intensif)


\begin{thebibliography}{}
\bibitem[Adams et al.(1989)]{1989ApJ...347..959A} Adams, F.~C., Ruden, S.~P., \& Shu, F.~H.\ 1989, ApJ, 347, 959 
\bibitem[Ataiee et al.(2013)]{2013A&A...553L...3A} Ataiee, S., Pinilla, P., Zsom, A., et al.\ 2013, A\&A, 553, L3
\bibitem[Bae et al.(2014)]{2014ApJ...795...61B} Bae, J., Hartmann, L., Zhu, Z., \& Nelson, R.~P.\ 2014, ApJ, 795, 61
\bibitem[Balbus \& Hawley(1991)]{1991ApJ...376..214B} Balbus, S.~A., \& Hawley, J.~F.\ 1991, ApJ, 376, 214
\bibitem[Baruteau \& Masset(2008)]{2008ApJ...678..483B} Baruteau, C., \& Masset, F.\ 2008, ApJ, 678, 483-497
\bibitem[Baruteau et al.(2011)]{2011MNRAS.416.1971B} Baruteau, C., Meru, F., \& Paardekooper, S.-J.\ 2011, MNRAS, 416, 1971 
\bibitem[Baruteau \& Papaloizou(2013)]{2013ApJ...778....7B} Baruteau, C., \& Papaloizou, J.~C.~B.\ 2013, ApJ, 778, 7 
\bibitem[Casassus et al.(2013)]{2013Natur.493..191C} Casassus, S., van der Plas, G., M, S.~P., et al.\ 2013, Nature, 493, 191
\bibitem[Chiang \& Laughlin(2013)]{2013MNRAS.431.3444C} Chiang, E., \& Laughlin, G.\ 2013, MNRAS, 431, 3444
\bibitem[Cossou et al.(2014)]{2014A&A...569A..56C} Cossou, C., Raymond, S.~N., Hersant, F., \& Pierens, A.\ 2014, A\& A, 569, A56
\bibitem[Crida et al.(2006)]{2006Icar..181..587C} Crida, A., Morbidelli, A., \& Masset, F.\ 2006, Icarus, 181, 587
\bibitem[D'Angelo et al.(2003)]{2003ApJ...599..548D} D'Angelo, G., Henning, T., \& Kley, W.\ 2003, ApJ, 599, 548 
\bibitem[de Val-Borro et al.(2006)]{2006MNRAS.370..529D} de Val-Borro, M., Edgar, R.~G., Artymowicz, P., et al.\ 2006, MNRAS, 370, 529
\bibitem[Faure et al.(2015)]{2015A&A...573A.132F} Faure, J., Fromang, S., Latter, H., \& Meheut, H.\ 2015, A\& A, 573, A132
\bibitem[Flock et al.(2017)]{2017ApJ...835..230F} Flock, M., Fromang, S., Turner, N.~J., \& Benisty, M.\ 2017, ApJ, 835, 230
\bibitem[Fu et al.(2014)]{2014ApJ...788L..41F} Fu, W., Li, H., Lubow, S., \& Li, S.\ 2014, ApJL, 788, L41 
\bibitem[Fukagawa et al.(2013)]{2013PASJ...65L..14F} Fukagawa, M., Tsukagoshi, T., Momose, M., et al.\ 2013, PASJ, 65, L14
\bibitem[Gammie(1996)]{1996ApJ...457..355G} Gammie, C.~F.\ 1996, ApJ, 457, 355
\bibitem[Gammie(2001)]{2001ApJ...553..174G} Gammie, C.~F.\ 2001, ApJ, 553, 174 
\bibitem[Gressel et al.(2012)]{2012MNRAS.422.1140G} Gressel, O., Nelson, R.~P., \& Turner, N.~J.\ 2012, MNRAS, 422, 1140
\bibitem[Hansen \& Murray(2012)]{2012ApJ...751..158H} Hansen, B.~M.~S., \& Murray, N.\ 2012, ApJ, 751, 158
\bibitem[Izidoro et al.(2017)]{2017MNRAS.470.1750I} Izidoro, A., Ogihara, M., Raymond, S.~N., et al.\ 2017, MNRAS, 470, 1750
\bibitem[Kley \& Dirksen(2006)]{2006A&A...447..369K} Kley, W., \& Dirksen, G.\ 2006, A\& A, 447, 369
\bibitem[Latter \& Balbus(2012)]{2012MNRAS.424.1977L} Latter, H.~N., \& Balbus, S.\ 2012, MNRAS, 424, 1977 
\bibitem[Les \& Lin(2015)]{2015MNRAS.450.1503L} Les, R., \& Lin, M.-K.\ 2015, MNRAS, 450, 1503
\bibitem[Lesur \& Papaloizou(2009)]{2009A&A...498....1L} Lesur, G., \& Papaloizou, J.~C.~B.\ 2009, A\& A, 498, 1
\bibitem[Li et al.(2000)]{2000ApJ...533.1023L} Li, H., Finn, J.~M., Lovelace, R.~V.~E., \& Colgate, S.~A.\ 2000, ApJ, 533, 1023 
\bibitem[Li et al.(2001)]{2001ApJ...551..874L} Li, H., Colgate, S.~A., Wendroff, B., \& Liska, R.\ 2001, ApJ, 551, 874
\bibitem[Lin \& Papaloizou(2011)]{2011MNRAS.415.1426L} Lin, M.-K., \& Papaloizou, J.~C.~B.\ 2011, MNRAS, 415, 1426 
\bibitem[Lin(2012)]{2012MNRAS.426.3211L} Lin, M.-K.\ 2012, MNRAS, 426, 3211
\bibitem[Lissauer et al.(2011)]{2011ApJS..197....8L} Lissauer, J.~J., Ragozzine, D., Fabrycky, D.~C., et al.\ 2011, ApJS, 197, 8
\bibitem[Lobo Gomes et al.(2015)]{2015ApJ...810...94L} Lobo Gomes, A., Klahr, H., Uribe, A.~L., Pinilla, P., \& Surville, C.\ 2015, ApJ, 810, 94
\bibitem[Lovelace et al.(1999)]{1999ApJ...513..805L} Lovelace, R.~V.~E., Li, H., Colgate, S.~A., \& Nelson, A.~F.\ 1999, ApJ, 513, 805
\bibitem[Lovelace \& Hohlfeld(2013)]{2013MNRAS.429..529L} Lovelace, R.~V.~E., \& Hohlfeld, R.~G.\ 2013, MNRAS, 429, 529
\bibitem[Lovis et al.(2011)]{2011A&A...528A.112L} Lovis, C., S{\'e}gransan, D., Mayor, M., et al.\ 2011, A\& A, 528, A112
\bibitem[Martin \& Lubow(2011)]{2011ApJ...740L...6M} Martin, R.~G., \& Lubow, S.~H.\ 2011, ApJL, 740, L6 
\bibitem[Masset(2000)]{2000A&AS..141..165M} Masset, F.\ 2000, A\&AS, 141, 165
\bibitem[Miranda et al.(2016)]{2016MNRAS.457.1944M} Miranda, R., Lai, D., \& M{\'e}heut, H.\ 2016, MNRAS, 457, 1944
\bibitem[M{\"u}ller et al.(2012)]{2012A&A...541A.123M} M{\"u}ller, T.~W.~A., Kley, W., \& Meru, F.\ 2012, A\& A, 541, A123 
\bibitem[Ogihara \& Ida(2009)]{2009ApJ...699..824O} Ogihara, M., \& Ida, S.\ 2009, ApJ, 699, 824
\bibitem[Okuzumi \& Hirose(2011)]{2011ApJ...742...65O} Okuzumi, S., \& Hirose, S.\ 2011, ApJ, 742, 65
\bibitem[Paardekooper et al.(2010)]{2010ApJ...725..146P} Paardekooper, S.-J., Lesur, G., \& Papaloizou, J.~C.~B.\ 2010, ApJ, 725, 146 
\bibitem[Papaloizou(2005)]{2005A&A...432..757P} Papaloizou, J.~C.~B.\ 2005, A\& A, 432, 757
\bibitem[Petersen et al.(2007)]{2007ApJ...658.1252P} Petersen, M.~R., Stewart, G.~R., \& Julien, K.\ 2007, ApJ, 658, 1252
\bibitem[Pierens et al.(2011)]{2011A&A...531A...5P} Pierens, A., Baruteau, C., \& Hersant, F.\ 2011, A\& A, 531, A5
\bibitem[Pierens et al.(2012)]{2012MNRAS.427.1562P} Pierens, A., Baruteau, C., \& Hersant, F.\ 2012, MNRAS, 427, 1562
\bibitem[Raymond \& Cossou(2014)]{2014MNRAS.440L..11R} Raymond, S.~N., \& Cossou, C.\ 2014, MNRAS, 440, L11 
\bibitem[Reg{\'a}ly et al.(2012)]{2012MNRAS.419.1701R} Reg{\'a}ly, Z., Juh{\'a}sz, A., S{\'a}ndor, Z., \& Dullemond, C.~P.\ 2012, MNRAS, 419, 1701
\bibitem[Reg{\'a}ly \& Vorobyov(2017)]{2017MNRAS.471.2204R} Reg{\'a}ly, Z., \& Vorobyov, E.\ 2017, MNRAS, 471, 2204 
\bibitem[Rein(2012)]{2012MNRAS.427L..21R} Rein, H.\ 2012, MNRAS, 427, L21
\bibitem[Richard et al.(2013)]{2013A&A...559A..30R} Richard, S., Barge, P., \& Le Diz{\`e}s, S.\ 2013, A\&A, 559, A30 
\bibitem[Shakura \& Sunyaev(1973)]{1973A&A....24..337S} Shakura, N.~I., \& Sunyaev, R.~A.\ 1973, A\& A, 24, 337
\bibitem[Umebayashi \& Nakano(1988)]{1988PThPS..96..151U} Umebayashi, T., \& Nakano, T.\ 1988, Progress of Theoretical Physics Supplement, 96, 151 
\bibitem[Terquem \& Papaloizou(2007)]{2007ApJ...654.1110T} Terquem, C., \& Papaloizou, J.~C.~B.\ 2007, ApJ, 654, 1110
\bibitem[van der Marel et al.(2013)]{2013Sci...340.1199V} van der Marel, N., van Dishoeck, E.~F., Bruderer, S., et al.\ 2013, Science, 340, 1199
\bibitem[van Leer(1977)]{1977JCoPh..23..276V} van Leer, B.\ 1977, Journal 
of Computational Physics, 23, 276 
\bibitem[Ward(1997)]{1997Icar..126..261W} Ward, W.~R.\ 1997, Icarus, 126, 261
\bibitem[Yellin-Bergovoy et al.(2016)]{2016GApFD.110..274Y} Yellin-Bergovoy, R., Heifetz, E., \& Umurhan, O.~M.\ 2016, Geophysical and Astrophysical Fluid Dynamics, 110, 274
\bibitem[Zhu et al.(2010)]{2010ApJ...713.1143Z} Zhu, Z., Hartmann, L., \& Gammie, C.\ 2010, ApJ, 713, 1143
\bibitem[Zhu et al.(2012)]{2012ApJ...746..110Z} Zhu, Z., Hartmann, L., Nelson, R.~P., \& Gammie, C.~F.\ 2012, ApJ, 746, 110
\bibitem[Zhu et al.(2014)]{2014ApJ...785..122Z} Zhu, Z., Stone, J.~M., Rafikov, R.~R., \& Bai, X.-n.\ 2014, ApJ, 785, 122
\bibitem[Zhu \& Baruteau(2016)]{2016MNRAS.458.3918Z} Zhu, Z., \& Baruteau, C.\ 2016, MNRAS, 458, 3918
\end{thebibliography}
\end{document}